\documentclass[a4paper,11pt]{article}
\pdfoutput=1 

\usepackage{jheppub} 

\newcommand{\xsep}{x_{\rm sep}}
\newcommand{\Mpl}{{M_{\rm Pl}}}

\definecolor{darkgreen}{RGB}{0,200,0}

\usepackage{soul} 

\title{\bf Simulating the Universe(s) III: Observables for the full bubble collision spacetime}

\author[a,b]{Matthew C. Johnson,}
\author[c]{Carroll L. Wainwright,}
\author[c]{Anthony Aguirre,}
\author[d]{Hiranya V. Peiris,}

\affiliation[a]{Department of Physics and Astronomy, York University \\ Toronto, On, M3J 1P3, Canada}
\affiliation[b]{Perimeter Institute for Theoretical Physics \\ Waterloo, Ontario N2L 2Y5, Canada}
\affiliation[c]{SCIPP and Department of Physics, University of California \\ Santa Cruz, CA, 95064, USA}
\affiliation[d]{Department of Physics and Astronomy, University College London \\ London WC1E 6BT, U.K.}

\emailAdd{mjohnson@perimeterinstitute.ca}
\emailAdd{aguirre@scipp.ucsc.edu}
\emailAdd{h.peiris@ucl.ac.uk}

\abstract{This is the third paper in a series establishing a quantitative relation between inflationary scalar field potential landscapes and the relic perturbations left by the collision between bubbles produced during eternal inflation. We introduce a new method for computing cosmological observables from numerical relativity simulations of bubble collisions in one space and one time dimension. This method tiles comoving hypersurfaces with locally-perturbed Friedmann-Robertson-Walker coordinate patches. The method extends previous work, which was limited to the spacetime region just inside the future light cone of the collision, and allows us to explore the full bubble-collision spacetime. We validate our new methods against previous work, and present a full set of predictions for the comoving curvature perturbation and local negative spatial curvature produced by identical and non-identical bubble collisions, in single scalar field models of eternal inflation. In both collision types, there is a non-zero contribution to the spatial curvature and cosmic microwave background quadrupole. Some collisions between non-identical bubbles excite wall modes, giving extra structure to the predicted temperature anisotropies. We comment on the implications of our results for future observational searches. For non-identical bubble collisions, we also find that the surfaces of constant field can readjust in the presence of a collision  to produce spatially infinite sections that become nearly homogeneous deep into the region affected by the collision. Contrary to previous assumptions, this is true even in the bubble into which the domain wall is accelerating.
}

\begin{document} 
\maketitle

\section{Introduction}

In the well-studied ``false vacuum" variant of eternal inflation, our universe is contained inside one bubble among many, each nucleated from a metastable false vacuum. The relic perturbations left by the collision between bubbles have been established as a quantitative observational probe of eternal inflation~\cite{Aguirre:2007an,Aguirre:2007wm,Aguirre:2008wy,Aguirre:2009ug,Feeney_etal:2010dd,Feeney_etal:2010jj,Feeney:2012hj,Johnson:2010bn,Johnson:2011wt,McEwen:2012uk,Wainwright:2013lea,Wainwright:2014pta,Zhang:2015uta,Chang_Kleban_Levi:2009,Chang:2007eq,Czech:2010rg,Freivogel_etal:2009it,Freivogel:2005vv,Gobbetti_Kleban:2012,Kleban_Levi_Sigurdson:2011,Kleban:2011pg,Salem:2012,Zhang:2015bga,Ahlqvist:2013whn,Amin:2013dqa,Amin:2013eqa,Easther:2009ft,Giblin:2010bd,Blanco-Pillado:2003hq,Deskins:2012tj,Freivogel:2007fx,Garriga:2006hw,Gott:1984ps,Hawking:1982ga,Hwang:2012pj,Kozaczuk:2012sx,Larjo:2009mt,Moss:1994pi,Osborne:2013hea,Osborne:2013jea,Wu:1984eda}. For a review of eternal inflation see Refs.~\cite{Aguirre:2007gy,Guth:2013epa}; for a big picture review of bubble collisions in eternal inflation, see Refs.~\cite{Aguirre:2009ug,Kleban:2011pg}. 

This paper is the third in a series which has established a quantitative connection between the scalar field Lagrangian underlying eternal inflation and cosmological observables~\cite{Wainwright:2013lea,Wainwright:2014pta}. The first two papers in the series served as an important proof of principle that bubble collisions can lead to {\em quantitative} constraints on the theory underlying eternal inflation. These predictions have been used to forecast the ability of near-term cosmic microwave background (CMB)~\cite{Feeney:2015hwa} and large scale structure~\cite{Zhang:2015uta} datasets to place constraints on the theory underlying eternal inflation. Previous work has already constrained the presence of bubble collisions using CMB data from the WMAP satellite~\cite{Feeney_etal:2010dd,Feeney_etal:2010jj,Feeney:2012hj,McEwen:2012uk,Osborne:2013hea,Osborne:2013jea}. Unfortunately, the method for extracting observables used in these previous works was limited to computing perturbations in the vicinity of the collision boundary. This prevented an assessment of observables far inside the spacetime region affected by the collision. In this paper, we overcome this limitation by introducing a new method for computing observables in a bubble collision spacetime that allows us to make predictions for {\em all} observers. This is accomplished by tiling the reheating surface with a continuous set of cosmological coordinate patches. Our new method allows us to address a number of outstanding questions regarding the overall structure of the collision spacetime and the possibility of new observables for bubble collisions in eternal inflation.

\section{Extracting cosmological observables for bubble collisions}\label{sec:extract}

To construct the bubble collision spacetimes, we use the simulation code described in Ref.~\cite{Wainwright:2013lea}. Within this framework, we assume the SO(2,1) symmetry of the collision space-time~\cite{Hawking:1982ga} between two Coleman-de Luccia vacuum bubbles~\cite{Coleman:1977py,Coleman:1980aw}, allowing us to perform simulations in one space and one time dimension. For the single-field models we study, this has been shown to be a good approximation in Ref.~\cite{Braden:2014cra}. 

Simulations are performed in the global foliation of the SO(2,1) symmetric collision space-time: 
\begin{equation}
\label{eq:metric1}
H_F^2 ds^2 = -\alpha(N,x)^2 dN^2 + a(N,x)^2 \cosh^2 N dx^2 + \sinh^2 N (d\chi^2 + \sinh^2\chi d\varphi^2),
\end{equation}
where $x$ is the simulation spatial variable (periodic with period $2\pi$), $N$ is a time variable which roughly measures the number of $e$-foldings in the surrounding eternally inflating de Sitter space,   $\alpha$ and $a$ are the simulated metric functions, and $H_F$ is the false vacuum Hubble constant. In the false vacuum outside the bubbles, i.e. in de Sitter space, $\alpha=a=1$ for all $N$. Generally, an observation bubble containing a phenomenologically viable epoch of inflation and a collision bubble (which may or may not contain the same vacuum), are input as initial conditions and then evolved using the coupled Einstein and scalar field equations. The initial conditions are fixed by nucleation physics~\cite{Coleman:1977py,Coleman:1980aw} and choosing an (arbitrary) reference frame, thus given a scalar field potential, the only free parameter in the simulation is the initial separation $\Delta \xsep$ between the colliding bubbles. The output is a complete description of the collision spacetime in terms of $\phi(N,x)$, $\alpha(N,x)$, and $a(N,x)$. In the following, we measure $\phi$ in terms of $M_{\rm Pl} \equiv G_N^{-1/2}$.

In order to make contact with observations, we define a coordinate system about each position in the simulation that corresponds to a perturbed Friedmann-Robertson-Walker (FRW) universe. The comoving gauge is most convenient, since the comoving curvature perturbation is conserved on superhorizon scales. In Ref.~\cite{Wainwright:2013lea} this was accomplished by evolving a set of geodesics through the simulation to construct a perturbed FRW universe in the synchronous gauge. A linear gauge transformation was then used to determine the comoving curvature perturbation from the metric perturbations in synchronous gauge. This method was rather limiting, since the extracted comoving curvature perturbation was only valid in the coordinate range where it was in the linear regime. In particular, it was impossible to probe deep within the collision region.

In this paper, we introduce a new method which allows us to directly extract the comoving curvature perturbation observed locally by any observer in the simulation. Conveniently, for single-field models, the comoving gauge is defined by slices of constant field. As we now outline, it is therefore possible to perform a coordinate transformation under which the induced metric on slices of constant field deriving from Eq.~\ref{eq:metric1} is explicitly the perturbed open FRW metric in comoving gauge.

\subsection{Computing the comoving curvature perturbation}

First, let us define a new variable $u$ that labels proper distance along the comoving slice (along which $N$ is not necessarily constant):
\begin{equation}
\label{eq:u}
u(x) = \int_0^x \sqrt{ (a \cosh N)^2 - \left(\alpha \frac{dN}{dx}\right)^2 } dx'.
\end{equation}
Note that $N$ is not a function of $x$; $dN/dx$ represents the change in $N$ with $x$ as one moves along the slice.
This results in the spatial metric
\begin{equation}
\label{eq:metric2}
H_F^2 ds^2 = du^2 + \sinh^2 N(u) (d\chi^2 + \sinh^2\chi d\varphi^2),
\end{equation}
where $N(u)$ is defined as value of $N$ obtained by moving the proper distance $u$ along the slice.

We then define a new variable $\xi$ that is a linear transformation of $u$,
\begin{equation}
\label{eq:uxi}
u - u_0 = a_0 (\xi - \xi_0)
\end{equation}
with the constants $u_0$, $a_0$ and $\xi_0$ as yet unspecified, such that Eq.~\ref{eq:metric2} can be written as
\begin{equation}
\label{eq:metric3}
H_F^2 ds^2 = a_0^2 [d\xi^2 + (1-2B)\cosh^2 \xi (d\chi^2 + \sinh^2\chi d\varphi^2)]
\end{equation}
where
\begin{equation}
\label{eq:B}
1-2B = \frac{\sinh^2 N(u)}{a_0^2 \cosh^2 \xi(u)}.
\end{equation}
When $B=0$, Eq.~\ref{eq:metric3} has constant negative spatial curvature, and represents a constant-time hypersurface in an anisotropic foliation of an open FRW universe (the anisotropic hyperbolic coordinates described in Ref.~\cite{Wainwright:2013lea}) with scale factor $a_0$. The metric function $B$ defines a scalar perturbation on top of an open FRW universe. 

We can fix the constants $\xi_0$ and $a_0$ by requiring that at an arbitrary position $u_0 \equiv u(x_0)$ on the slice we have
\begin{equation}
B(u_0) = 0, \ \ \ \frac{dB}{d\xi} (u_0) = 0 
\end{equation}
which yields
\begin{gather}
\label{eq:xi_0}
\xi_0 = \sinh^{-1}\left(\cosh N_0  \frac{d N_0}{du_0} \right) \\
\label{eq:a_0}
a_0 = \frac{\sinh N_0}{\cosh \xi_0},
\end{gather}
where $N_0 = N(u_0)$. The choice $B(u_0) = 0$ amounts to absorbing the local expansion into the scale factor $a_0$. The choice $dB/d\xi=0$ corresponds to requiring the observer to be at rest with respect to the spatial slice (gradients of $B$ would induce a peculiar velocity). The constant $\xi_0$ corresponds to the observer's position in the anisotropic hyperbolic coordinates.

We now find the Ricci 3-scalar as a function of the perturbation $B$ (neglecting the conformal factor $a_0$),
\begin{equation}
\begin{split}
\label{eq:ricci}
R^{(3)} =& -6 - 32 B^2 (B^2-1) 
+\frac{4B}{\cosh^2\xi} (8B^3 + 4B^2 -2B-1) \\ 
&- 4\tanh\xi (  8B^3 + 4B^2- 10B - 3 ) \partial_\xi B \\
&-2 (4B^2+4B-1) (\partial_\xi B)^2
+ 4 (2B+1) \partial^2_\xi B.
\end{split}
\end{equation}
Note that $R^{(3)}=-6$ when $B=0$, representing the overall negative spatial curvature of comoving slices in unperturbed bubbles. The Ricci scalar can also be written in terms of the scalar curvature perturbation $\mathcal{R}$,
\begin{equation}\label{eq:R}
R^{(3)}(\xi) = -6 + 4 \nabla^2\mathcal{R}.
\end{equation}
Since $R^{(3)}$ is a function of $\xi$ only, the perturbation $\mathcal{R}$ depends only upon $\xi$ as well. The Laplacian is then given by
\begin{equation}
\label{eq:psi}
\nabla^2\mathcal{R} = \left(\partial_\xi^2 + \frac{2\tanh\xi - 2\partial_\xi B}{1-2B}\partial_\xi\right)\mathcal{R}.
\end{equation}
Plugging Eq.~\ref{eq:psi} into the right-hand side of Eq.~\ref{eq:R}, and Eq.~\ref{eq:ricci} into the left side gives an elliptic differential equation for $\mathcal{R}$ in terms of $B$. This can then be integrated to find $\mathcal{R}(\xi)$, with the integration constant fixed by $\mathcal{R} = d\mathcal{R}/d\xi = 0$ at the observer position $\xi_0$ (this corresponds to an observer at rest with respect to the spatial hypersurface).

Each position $x_0$ on a slice of constant field defines a perturbed open FRW coordinate patch characterized by a scale factor $a_0 (x_0)$ and curvature perturbation $\mathcal{R}(\xi | x_0)$. We cover the entire constant field spatial slice by the set of all such patches. The set of patches characterize observables at each comoving position after inflation in the collision space-time.

\subsection{Cartesian coordinates}

Observables are most easily computed by going to Cartesian coordinates $X,Y,Z$ and translating an observer at some fiducial point $\xi_0$ to the origin, where the metric is given by
\begin{equation}\label{eq:unperturbed_cartesian}
H_F^2 ds^2 = -d\tau^2 + \left[\frac{a(\tau)}{1-\tfrac{R^2}{4}}\right]^2\left(dX^2 + dY^2 + dZ^2\right) .
\end{equation}
This is convenient because  the past light cones of observers projected onto a constant time hypersurface are spheres centered on the origin of hyperbolic Cartesian coordinates. The metric for anisotropic hyperbolic coordinates is given by 
\begin{equation}\label{eq:anisotropic_c}
H_F^2 ds^2 = -d\tau^2 + a(\tau)^2 \left[ d\xi^2 + \cosh^2 \xi \left( d\rho^2 + \sinh^2 \rho d\varphi^2 \right) \right] .
\end{equation}
Finding the cartesian coordinates in terms of the anisotropic hyperbolic coordinates, we obtain:
\begin{align}
\label{eq:xiToX}
X &= \frac{2  \sinh\xi}{1+\cosh\xi \cosh\rho} \\
\label{eq:Y_from_xirho}
Y &= \frac{2 \sinh\rho\cosh\xi}{1+\cosh\xi\cosh\rho} \cos \varphi \\
Z &= \frac{2 \sinh\rho\cosh\xi}{1+\cosh\xi\cosh\rho} \sin \varphi,
\end{align}
or equivalently, 
\begin{align}\label{eq:xirhophi}
\sinh\xi &= \frac{X}{\left( 1-\frac{R^2}{4} \right)},&\;\; 
\tanh\rho &= \frac{\sqrt{Y^2+Z^2}}{ \left( 1+\frac{R^2}{4} \right)},&\;\;
\tan\varphi &= \frac{Z}{Y}.
\end{align}

The symmetry of the collision spacetime implies that different observers on the same constant $\xi$ slice are equivalent. We can therefore take any observer's position to be at $\rho=0$. To move to a frame where the observer is at the origin of Cartesian coordinates, we perform the following coordinate transformation:
\begin{eqnarray}
\label{eq:xi_boost}
\sinh \xi' &=& \cosh \xi_0 \sinh \xi  - \sinh \xi_0 \cosh \xi \cosh \rho, \\
\label{eq:rho_boost}
\cosh \rho' &=& \frac{\cosh \xi_0 \cosh \xi \cosh \rho - \sinh \xi_0 \sinh \xi}{\cosh \xi'}, \\
\varphi' &=& \varphi .
\end{eqnarray}
Along $\rho = 0$ this corresponds to a translation of a point at $\xi = \xi_0$ to $\xi = 0$. For more details on these coordinate transformations, see Ref.~\cite{Wainwright:2013lea}. 

\begin{figure}[h]
   \centering
   \includegraphics[width=6cm]{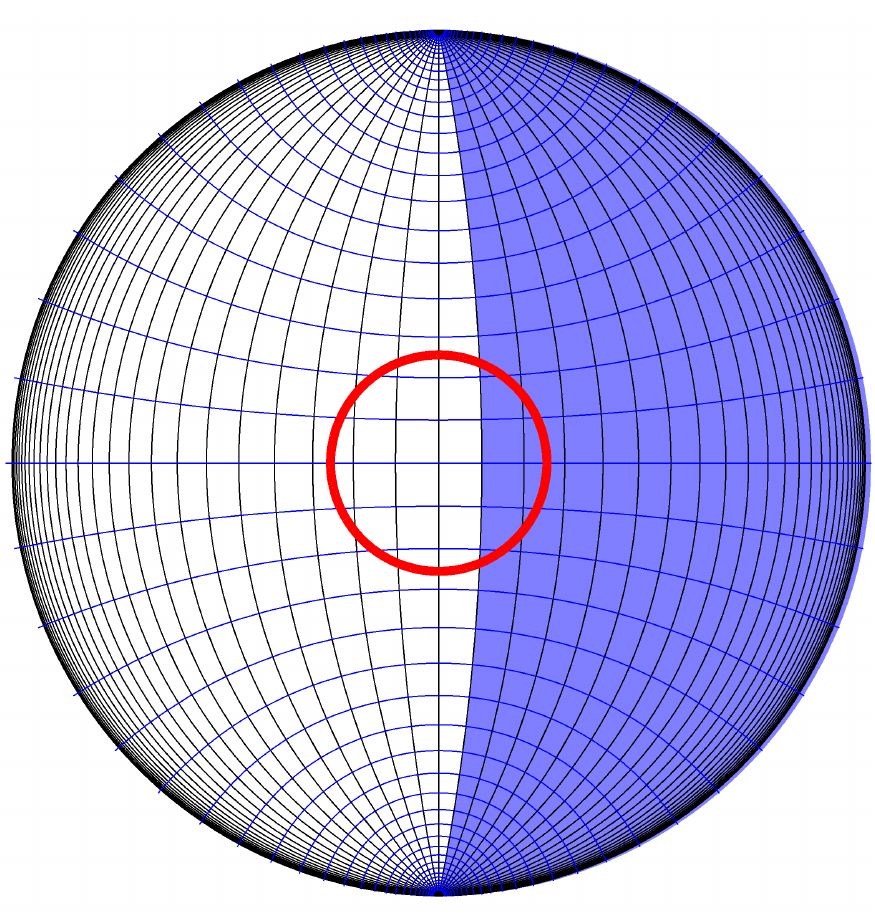}
   \includegraphics[width=6cm]{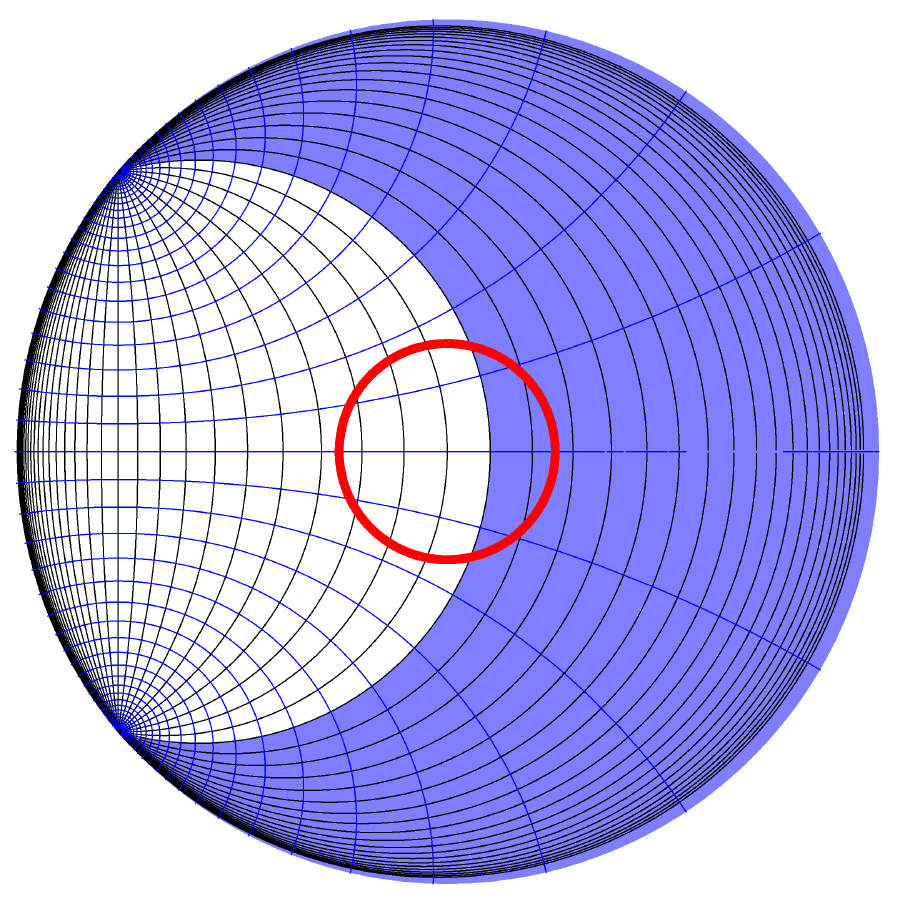}
   \caption{Constant FRW time hypersurfaces in the post-collision spacetime, represented in terms of the hyperbolic Cartesian coordinates. The hyperbolic Cartesian coordinates are bounded between $0< R < 2$; the edge of the disc is an infinite proper distance from the center. Vertical solid lines correspond to surfaces of constant $\xi$ and horizontal solid lines surfaces of constant $\rho$ in the anisotropic hyperbolic foliation of open FRW. The symmetries of the collision spacetime imply that the collision is independent of $\rho$. In the left panel, we show a collision spacetime for the reference point $\xi_0=0$. In the right panel, we show a collision spacetime with the reference point at $\xi_0 = 1$ after a translation in $\xi$ that brings the reference point to the origin of Cartesian coordinates. In both cases, the shaded region encloses the future light cone of a collision, and the red circle corresponds to the projection of the past light cone of a hypothetical observer at the origin. For observers at large $\xi_0$, the boundary of the future light cone becomes somewhat curved.}
   \label{fig:poincare}
\end{figure}

The observer position $\xi_0$ is an output of the algorithm described in the previous section. The main implication of the observer position is the shape of the surfaces of constant $\mathcal{R}$ in Cartesian coordinates. This is illustrated in Fig.~\ref{fig:poincare}, which shows a constant-time hypersurface in the Poincare disc representation of an open universe. In the left panel, $\xi_0 = 0$, and the future light cone of a collision (the boundary is located at $\xi=0.2$) is enclosed in the shaded region. $\mathcal{R}$ is constant on lines of constant-$\xi$ (black, vertical lines in the figure), and the projection of a past light cone is denoted by the red circle (the observable portion of the universe lying within the circle). The right panel shows the result for an observer located at $\xi_0 = 1.0$ who is translated to the origin of Cartesian coordinates, as well as the future light cone of a collision whose boundary is located at $\xi=1.2$. The distance in $\xi$ between the observer and collision is identical in both cases, but the observer at $\xi_0 = 1.0$ sees surfaces of constant $\mathcal{R}$ that are somewhat curved. The extent to which this surface looks curved is determined both by the observer's position and the size of his or her past light cone.

We quantify the departure from planar symmetry by computing the value of $\xi$ along an observer's light cone at a constant time with radius $R$. With $\theta$ as the viewing angle along the cone (by symmetry, there are no variations along the azimuthal angle), we find
\begin{equation}
\xi = \sinh^{-1} \left(\frac{4 R \cosh \xi_0 \cos \theta  + \left( 4+R^2  \right) \sinh \xi_0 }{4-R^2}\right)\, .
\end{equation}
In the limit where $R$ and $\xi-\xi_0$ are small, we recover $\xi - \xi_{0} \simeq R \cos \theta$. In the same limit, we see from Eq.~\ref{eq:xirhophi} that $\xi-\xi_0 \sim X$ near the origin, another way of highlighting the apparently planar symmetry of the collision.

\subsection{CMB observables}\label{sec:CMB_observables}

\begin{figure}[h]
   \centering
   \includegraphics[width=12cm]{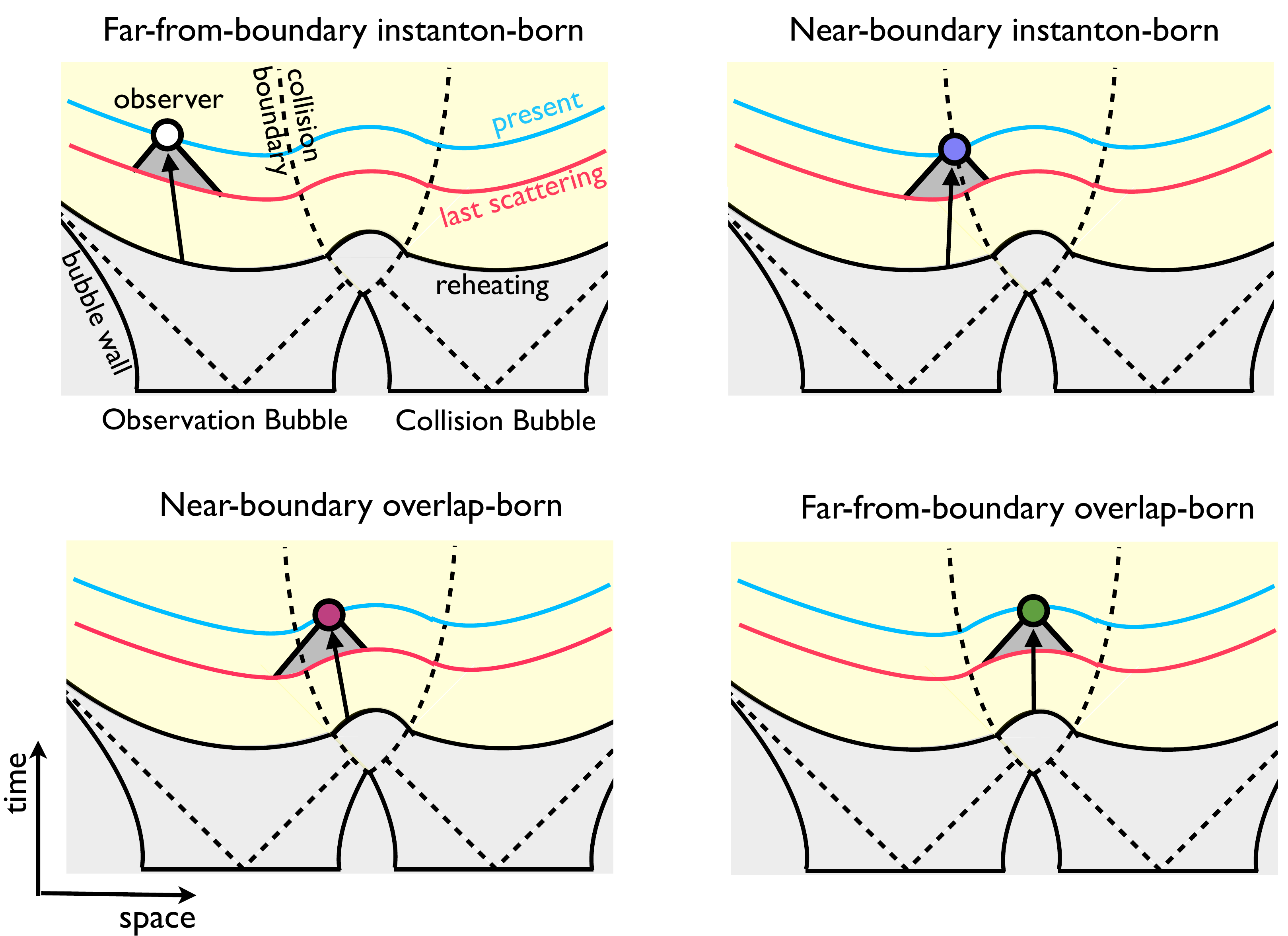}
   \caption{Spacetime diagrams depicting the collision between two identical bubbles: an observation bubble that houses a set of hypothetical observers, and a collision bubble. The collision boundary is the future light cone of the collision, which separates each bubble interior into the collision region (the causal future of the collision) and a region unaffected by the collision. Hypothetical observers originate on the reheating surface and follow comoving geodesics until the present (arrows). Each hypothetical observer has causal access to a different part of the surface of last scattering (shaded past light cones). Far-from-boundary instanton-born observers (top left) originate outside the collision region and do not have causal access to the collision boundary at last scattering. Near-bounary instanton-born observers (top right) originate outside the collision region, but have causal access to the collision boundary at last scattering. Near-boundary overlap-born observers (bottom left) originate from the collision region, and have causal access to the collision boundary at last scattering. Far-from-boundary overlap-born observers (bottom right) originate from the collision region, and do not have causal access to the collision boundary at last scattering.}
   \label{fig:causalpicture}
\end{figure}

\begin{figure}[h]
   \centering
   \includegraphics[width=12cm]{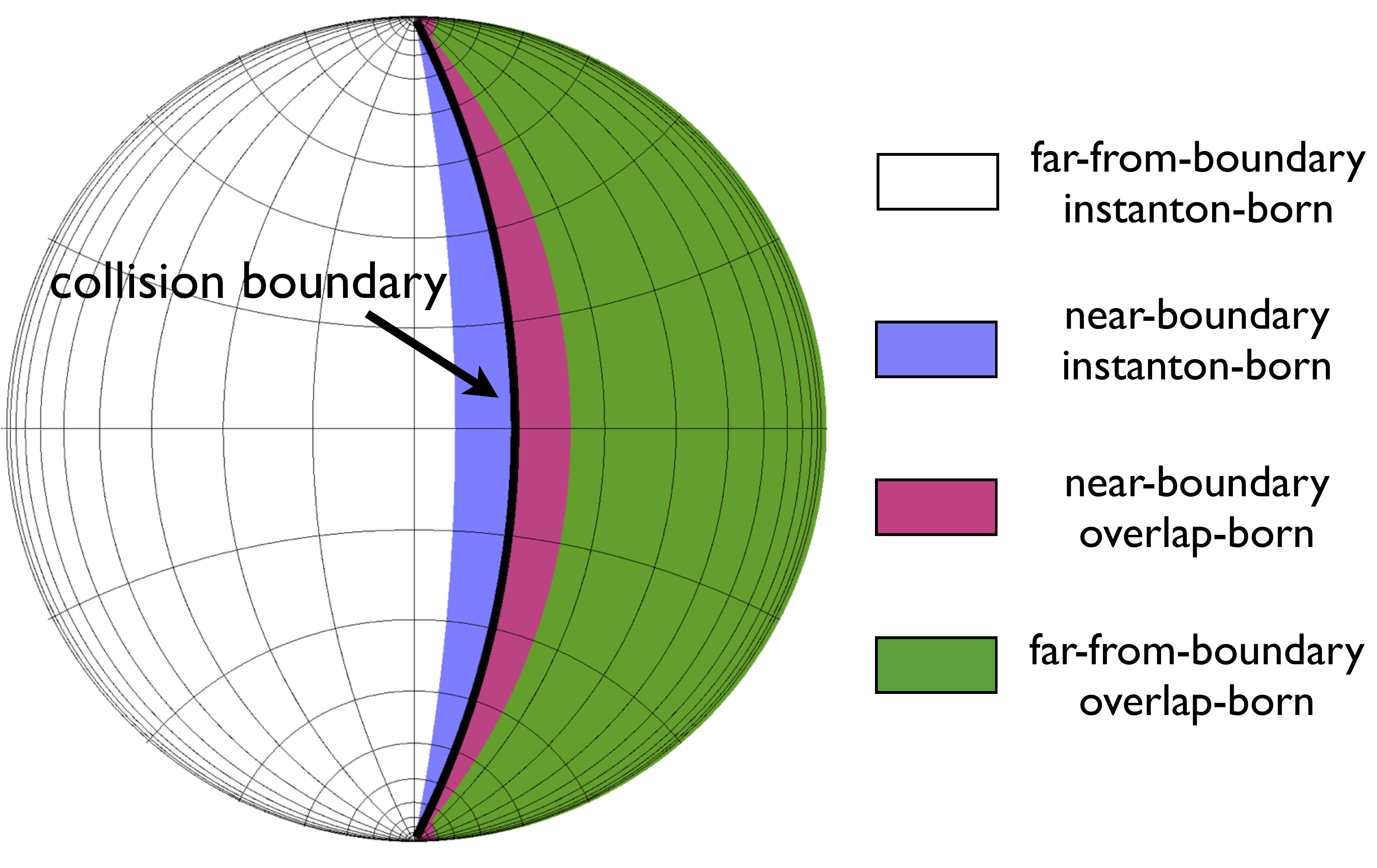}
   \caption{A constant FRW time hypersurface in the post-collision spacetime, represented in terms of the hyperbolic Cartesian coordinates. This diagram depicts the observation-side of the collision; an identical diagram could represent the collision-side. Shading represents the various regions that contain qualitatively different classes of observers. The thick solid line represents the collision boundary, which separates instanton-born from overlap-born observers. Both classes of observers can be near-boundary, where the observers have causal access to the collision boundary, or far-from-boundary, where observers do not have causal access to the collision boundary.}
   \label{fig:observer_names}
\end{figure}

In the absence of any bubble collisions, each bubble contains an infinite open FRW Universe that undergoes an epoch of slow-roll inflation, followed by reheating and standard cosmological evolution. We imagine that each bubble is populated by a set of hypothetical observers, each of whom (by homogeneity and isotropy) would make identical observations. However, once collisions are taken into account, the Universe inside each bubble is no longer homogeneous or isotropic, giving rise to classes of observers who would make qualitatively different observations depending on their position. Before proceeding, it is therefore useful to outline a set of terminology that will assist us in describing the various qualitatively different regions in the collision space-time and the associated hypothetical observers in each. Summary figures describing our terminology for the collision spaceime and associated observers are shown in Fig.~\ref{fig:causalpicture} and Fig.~\ref{fig:observer_names}. 

Fig.~\ref{fig:causalpicture} depicts 4 copies of a space-time diagram for a collision between two identical bubbles, highlighting the various types of observers. One bubble is denoted as the ``observation bubble," and the other the ``collision bubble." In the case of the collision between identical bubbles, whose interiors simply merge, there is no meaningful distinction between the two. In the case of collisions between non-identical bubbles, whose interiors are separated by a domain wall, the distinction is more important, with the observation bubble containing the hypothetical observers that we wish to describe and the collision bubble acting to perturb the observation bubble interior. To the extent that the bubble walls are compact, the collision is an event. The future light cone of the collision, the ``collision boundary," splits the observation bubble into two regions: one that is outside the causal future of the collision, and one that is inside the causal future of the collision. We denote the latter as the ``collision region."    

As depicted in Fig.~\ref{fig:causalpicture}, our proxy for an observer is a comoving geodesic originating on the reheating surface and terminating at the present (e.g. defined by a constant-density hypersurface). Each such observer has causal access to a different part of the surface of last scattering, shown as the shaded past light cones in Fig.~\ref{fig:causalpicture}. Figure~\ref{fig:observer_names} shows a constant-time hypersurface in hyperbolic Cartesian (comoving) coordinates, representing, for example, the surface of last scattering. An observer type is specified by the region of the reheating surface from which they were ``born" and which region of the surface of last scattering they now have causal access to. The various types of observers for identical bubble collisions are labeled in Figs.~\ref{fig:causalpicture} and~\ref{fig:observer_names}.

To delineate the various observer types, we denote by ``instanton-born" observers those that originate in the undisturbed observation or collision bubble. An observer who does not have causal access to the collision boundary on the surface of last scattering is ``far-from-boundary," while an observer that does have causal access to the collision boundary on the surface of last scattering is ``near-boundary." The ``Overlap-born" observers originate in the collision region.  They may be near-boundary observers (so that they have causal access to the collision boundary at last scattering) or ``far-from-boundary" (where they never see either undisturbed bubble).  The far-from-boundary overlap-born observers may, for the case of non-identical bubbles, be either on the observation bubble side (``observation-side") or collision bubble side (``collision-side") of the domain wall between the bubbles.  Thus, for example, previous papers in this series treated near-boundary instanton-born observation-side observers, but were not able to treat the  near-boundary overlap-born observation-side observers that originate just inside, rather than just outside, the collision boundary. 

Each observer position defines a local open universe of some curvature with a nearly planar-symmetric curvature perturbation. The locally-observed energy density in curvature $\Omega_k$ is related to the total expansion of the universe during inflation. Because collisions change the cosmological evolution, there is a potentially large variation in $\Omega_k$. Instanton-born observers all experience the same curvature $\Omega_k^{(\rm I)}$, which is determined by the number of $e$-folds in the unperturbed portion of the collision space-time and the (unspecified) details of reheating. Overlap-born observers at different positions $x_0$ on the same constant-field surface experience different curvatures $\Omega_k^{\rm (O)} (x_0)$, given by
\begin{equation}\label{eq:local_curvature}
\Omega_k^{\rm (O)} = \Omega_k^{(\rm I)} \left(\frac{a_{(\rm I)}}{a_0 (x_0)}\right)^2 \, , 
\end{equation}
where $a_{(\rm I)}$ is the scale factor for instanton-born observers at the end of inflation and $a_0 (x_0)$ is the scale factor at the end of inflation at position $x_0$.

Cosmological observables are determined by the comoving curvature perturbation at the end of inflation $\mathcal{R}(\xi | x_0)$, which is a function of position $x_0$. In this paper, we will primarily be interested in the imprint of bubble collisions on the CMB temperature anisotropies. Given the comoving curvature perturbation as an input, the temperature and polarization anisotropies are most accurately computed using a standard Boltzmann code such as CAMB; this has been implemented in Refs.~\cite{Czech:2010rg,Kleban_Levi_Sigurdson:2011,Feeney:2015hwa} for near-boundary instanton-born observers. Predictions for the CMB signature seen by near-boundary overlap-born observers are most accurately computed in the same fashion. However, for far-from-boundary overlap-born observers (whose past light cones do not intersect the collision boundary) we can obtain an accurate estimate for the temperature anisotropy in the Sachs Wolfe approximation.

For far-from-boundary overlap-born observers, the locally observed comoving curvature perturbation is continuous in all its derivatives. We can therefore perform a Taylor series expansion about an observer at $\xi_0$ to obtain
 \begin{equation}
\mathcal{R}(\xi | x_0) \simeq \frac{1}{2} \partial_\xi^2 \mathcal{R}(\xi_0|x_0) \ (\xi-\xi_0)^2 + \frac{1}{3} \partial_\xi^3 \mathcal{R}(\xi_0|x_0) \ (\xi-\xi_0)^3 + \ldots,
\end{equation}
where the constant and linear terms are zero by definition. For small curvature, we can use $\xi-\xi_0 \simeq R \cos \theta$ to obtain
 \begin{equation}
\mathcal{R} (\xi|x_0)\simeq \frac{1}{2} \partial_\xi^2 \mathcal{R}(\xi_0|x_0) \ R^2 \cos^2 \theta + \frac{1}{3} \partial_\xi^3 \mathcal{R}(\xi_0|x_0) \ R^3 \cos^3 \theta + \ldots
\end{equation}

In the Sachs-Wolfe approximation, the CMB temperature anisotropies are related to the projected comoving curvature on the surface of last scattering by
\begin{equation}\label{eq:dtsachswolfe}
\frac{\Delta T}{T} \simeq \frac{\mathcal{R} (R_{\rm ls}|x_0)}{5}.
\end{equation}
The distance to the surface of last scattering depends on the local value of the curvature, given by
\begin{equation}
R_{\rm ls} (x_0) = 2 \sqrt{\Omega_k^{\rm (O)}} .
\end{equation}
Going to harmonic space
\begin{equation}
\frac{\Delta T}{T} = \sum_{\ell, m} a_{\ell m} Y_{\ell, m} (\theta, \varphi) \, ,
\end{equation} 
and identifying powers of $\cos \theta$ with the spherical harmonics, the temperature quadrupole is given by
\begin{eqnarray}\label{eq:multipoles}
a_{20} &=&  \Omega_k^{\rm (O)} \ \frac{8 \sqrt{\pi}}{15 \sqrt{5}} \ \partial_\xi^2 \mathcal{R}(\xi_0|x_0) \, .
\end{eqnarray}
For higher $\ell$, spherical harmonic coefficients are suppressed by $a_{\ell 0} \propto \left( \Omega_k^{\rm (O)}\right)^{\ell/2} \ \partial_\xi^\ell \mathcal{R}(\xi_0|x_0)$. Thus unless higher derivatives of the comoving curvature grow faster than this curvature suppression, all but the lowest multipoles will be negligible in the phenomenologically relevant low-curvature regime. For far-from-boundary overlap-born observers, the leading observables are therefore $\Omega_k^{\rm (O)}$ and $a_{20}$, both of which are  determined in the simulation for observers at all possible vantage points. 

\section{Numerical Implementation}

We have numerically implemented the above procedure for computing cosmological observables using the output of the simulation code introduced in Ref.~\cite{Wainwright:2013lea}. We concentrate on single-field models below.

For single-field models, a spatial slice is defined by $\phi(x,N) =  \phi_0$ for some constant $\phi_0$. Any choice of $\phi_0$ that is sufficiently far down the inflationary plateau for the comoving curvature perturbation to freeze in is acceptable. We use the Brent method for root finding to numerically calculate  $N(x)$ along the slice. Eq.~\ref{eq:u} is numerically integrated to obtain $u(x)$, and the result used to compute $\xi$ via Eq.~\ref{eq:uxi} defined about a point $x_0$. With these quantities, we compute $B$ using Eq.~\ref{eq:B}, and with $B$ the three-curvature Eq.~\ref{eq:ricci}. The local scale factor $a_0$ is computed from Eq.~\ref{eq:a_0}. Comparing this to the scale factor in the unperturbed portion of the bubble yields $\Omega_k^{\rm (O)} / \Omega_k^{\rm (I)}$ from Eq.~\ref{eq:local_curvature}. Finally, we integrate Eq.~\ref{eq:psi} to obtain the comoving curvature perturbation $\mathcal{R}(\xi | x_0)$, which can be converted to multipoles using Eq.~\ref{eq:multipoles}. This procedure is repeated for all points $x_0$ on the constant-field slice.

\section{Simulating collisions between identical bubbles}\label{sec:identical}

We begin by considering collisions between identical bubbles. In this case, each bubble contains the same true vacuum, and the bubble interiors merge. This gives rise to a  spacelike hypersurface at late times that encompasses the interior of both bubbles. We cover late time spacelike hypersurfaces by a set of cosmological patches using the method outlined in Sec.~\ref{sec:extract}. 

\begin{figure}
   \includegraphics{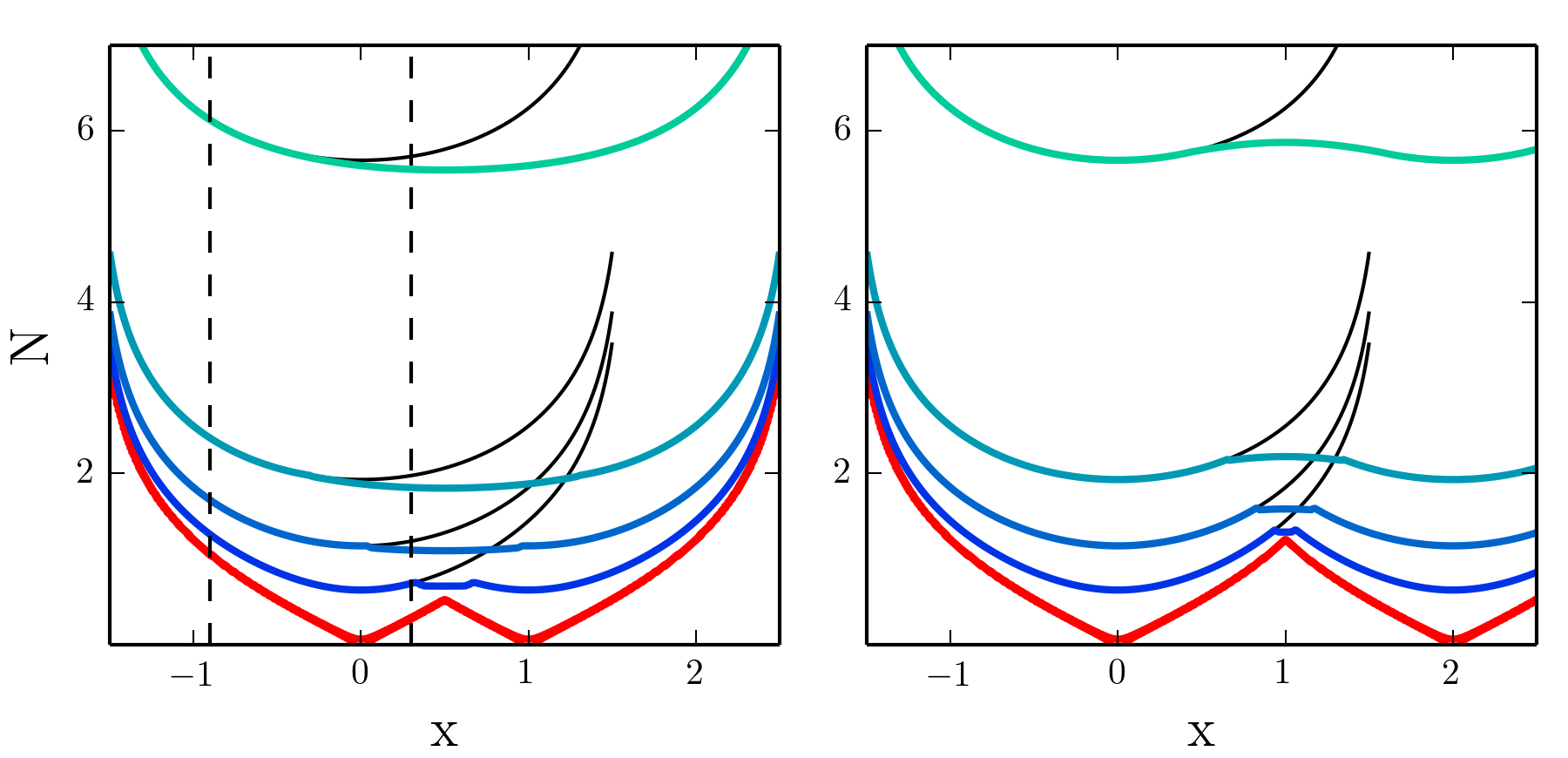}
   \caption{Contour plots of $\phi(x, N)$ for the collision between two identical bubbles for initial separations $\Delta \xsep =1$ (left panel) and $\Delta \xsep =2$ (right panel). The red contour ($\phi = 10^{-4}\, \Mpl$) tracks the bubble wall, successive hypersurfaces ($\phi = 0.005, 0.013, 0.03, 0.13\, \Mpl$) are shown with blue to green contours, and the corresponding contours in an unperturbed spacetime are shown in grey. Vertical dashed lines denote the position of two sample observers, one near-boundary instanton-born observer and one far-from-boundary overlap-born observer. The time evolution of the comoving curvature perturbation experienced by these two observers as a function of the anisotropic hyperbolic cosmological coordinates is shown in Fig.~\ref{fig:timedependence}.
}
   \label{fig:identical_observers_contours}
\end{figure}

We perform a simulation of identical colliding bubbles using the ``quartic barrier" potential of Ref.~\cite{Wainwright:2014pta} with parameters $\mu=0.01$, $\omega=0.4$, $\Delta\phi=8\times 10^{-4}\, \Mpl$, $\phi_0=3\, \Mpl$, and $\Delta \xsep = 1, 2$. Contour plots depicting $\phi(x, N)$ in the collision spacetime are shown in Fig.~\ref{fig:identical_observers_contours} for initial separations of $\Delta \xsep =1$ (left panel) and $\Delta \xsep =2$ (right panel). Contours for $\phi = 10^{-4}\, \Mpl$ (red) and $\phi = 0.005, 0.013, 0.03, 0.13\, \Mpl$ (blue to green) are shown, encompassing roughly 5 $e$-folds of inflation. At late times, it can be seen that a continuous  hypersurface encompassing the interiors of both bubbles is formed. For reference, surfaces of constant field in an unperturbed bubble (grey) are plotted as well. 

\begin{figure}
   \centering
   \includegraphics{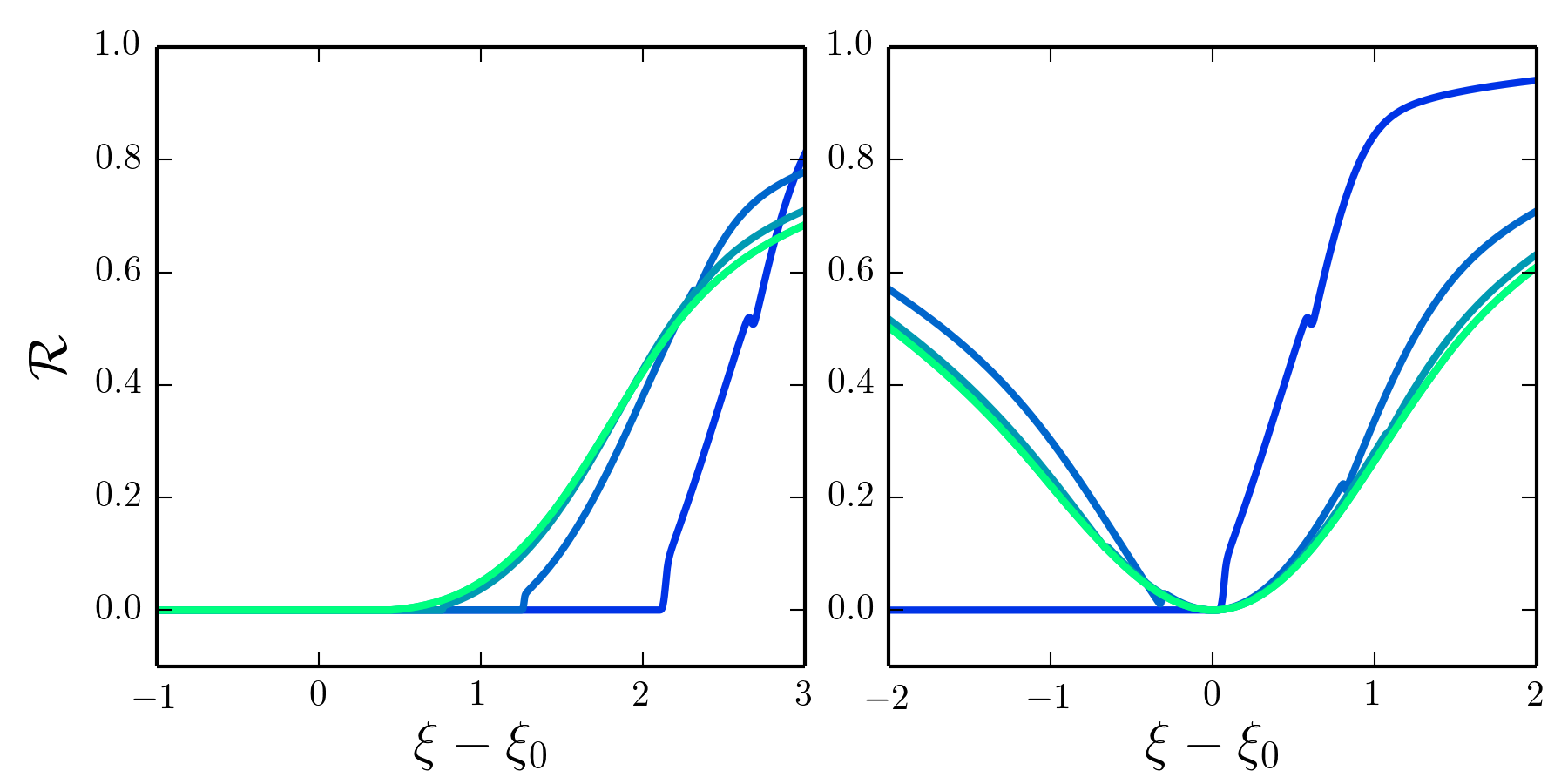}
   \caption{$\mathcal{R} (\xi)$ on slices of $\phi = 0.005, 0.013, 0.03, 0.13\, \Mpl$ (blue to green) as seen by observers at $x_0 = -0.9$ (left panel) and $x_0 = 0.3$ (right panel) for the collision depicted in the left panel  of Fig.~\ref{fig:identical_observers_contours} with $\Delta \xsep =1$. The comoving curvature perturbation propagates into the observation bubble (right to left in the figure), freezing in after approximately 5 $e$-folds.}
   \label{fig:timedependence}
\end{figure}

In Fig.~\ref{fig:identical_observers_contours}, we see that the perturbation from the collision propagates into each bubble, travelling a fixed comoving distance. We re-cast this time evolution in terms of the anisotropic hyperbolic coordinates surrounding a pair of observers in Fig.~\ref{fig:timedependence}, where we plot $\mathcal{R}$ for the collision with $\Delta \xsep = 1$ as a function of $\xi$ on slices of $\phi = 0.005, 0.013, 0.03, 0.13\, \Mpl$ (blue to green), matching the contours in Fig.~\ref{fig:identical_observers_contours}. The left panel is centered on $x_0 = -0.9$, and the right panel is centered on $x_0 = 0.3$; these positions are denoted by the vertical dashed lines in the left panel of Fig.~\ref{fig:identical_observers_contours}. In both cases, the perturbation propagates into the observation bubble, quickly converging to a constant amplitude and position. As expected, the comoving curvature perturbation is frozen in as inflation progresses. Because $x_0=-0.9$ lies outside the collision boundary, this vantage is inhabited by an near-boundary instanton-born observer. An observer at this position would have causal access to regions that are affected by the collision and regions that are not. The position $x_0 = 0.3$ corresponds to a far-from-boundary overlap-born observer, who experiences a curvature perturbation that is non-zero everywhere and nearly quadratic. 

In the unperturbed portion of the bubble, the scale factor evolves as $a_0 \simeq {\rm sinh} (H_I t)$. Defining the number of $e$-folds as $N_e = {\rm arcsinh} \ a_0$, the time slices shown in Fig.~\ref{fig:timedependence} correspond to $N_e = 0.5, 1.0, 2.0, 5.0$ $e$-folds in the unperturbed portion of the bubble. In this example, the curvature perturbation freezes in after approximately $5$ $e$-folds, corresponding to $\phi_0 \sim 0.13\, \Mpl$. Therefore, choosing $\phi_0 > 0.13\, \Mpl$ will give a valid representation of the late-time comoving curvature perturbation. All models examined in this paper utilize the same slow-roll potential, allowing us to make a single choice for the constant $\phi$ hypersurface, which we set to be $\phi_0 = 0.3\, \Mpl$. 

\begin{figure}
   \centering
   \includegraphics{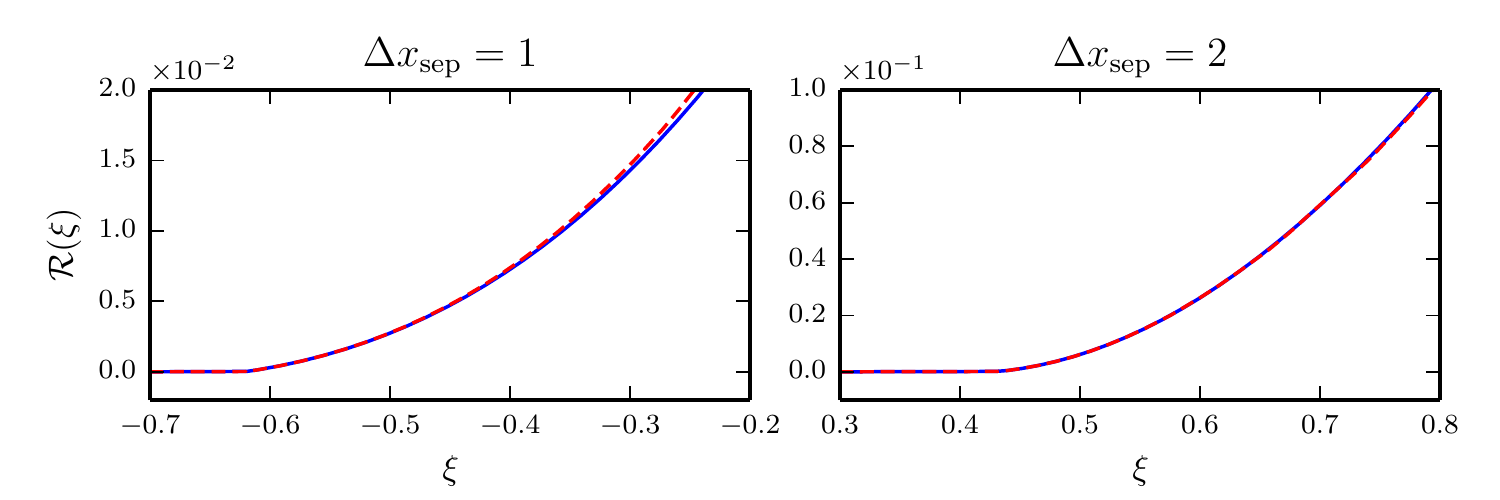}
   \caption{Comparing the comoving curvature perturbation calculated using the older geodesic method (dashed red lines) to the calculation with the new comoving method (solid blue lines).}
   \label{fig:geo_comparison}
\end{figure}

As a check of our method, in Fig.~\ref{fig:geo_comparison} we compare the comoving curvature perturbation obtained using the new method introduced in this paper with the results for $\mathcal{R}(\xi | x_0)$ obtained using the geodesic method of Ref.~\cite{Wainwright:2013lea}. Recall that the geodesic method is only valid for instanton-born observers,  so we choose a reference point $x_0$ outside the collision region. Very near the collision boundary, the two methods display excellent agreement.  Further from the collision boundary, there is some visible disagreement between the two curves (at the percent level). This can be accounted for by considering two effects: the definition of $\xi$ at a given spacetime point differs slightly between the two methods, and the geodesic method relies on small slow-roll parameters to transform from synchronous to comoving gauge.

\begin{figure}
   \centering
   \includegraphics{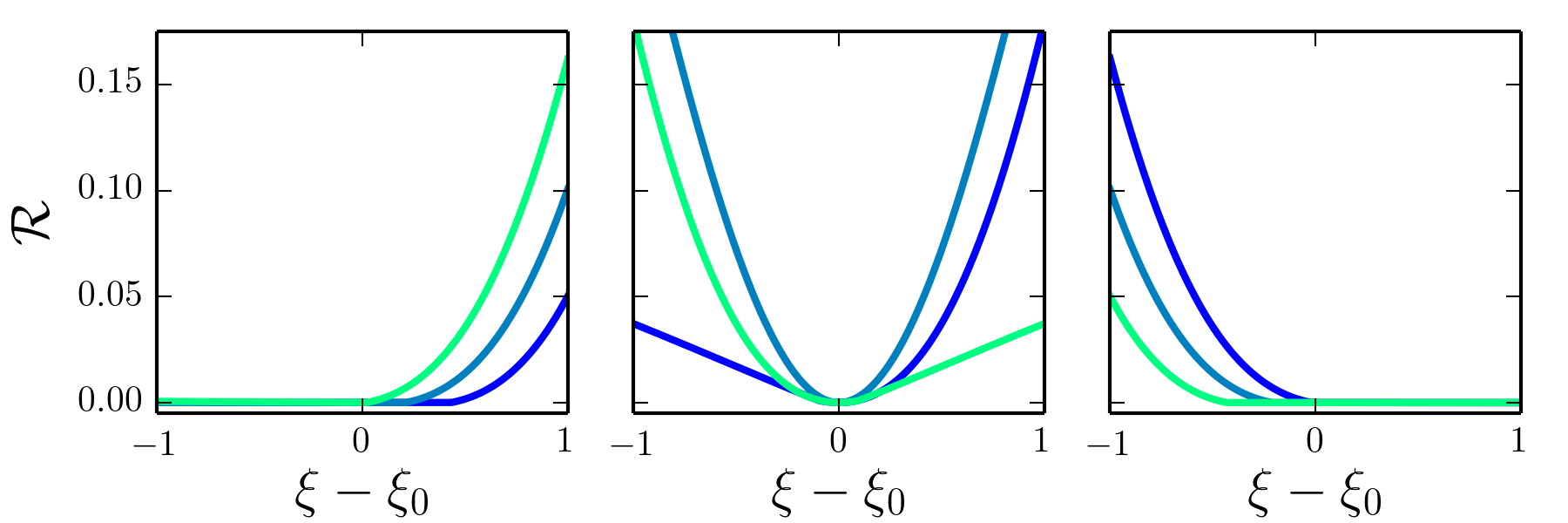}
   \includegraphics{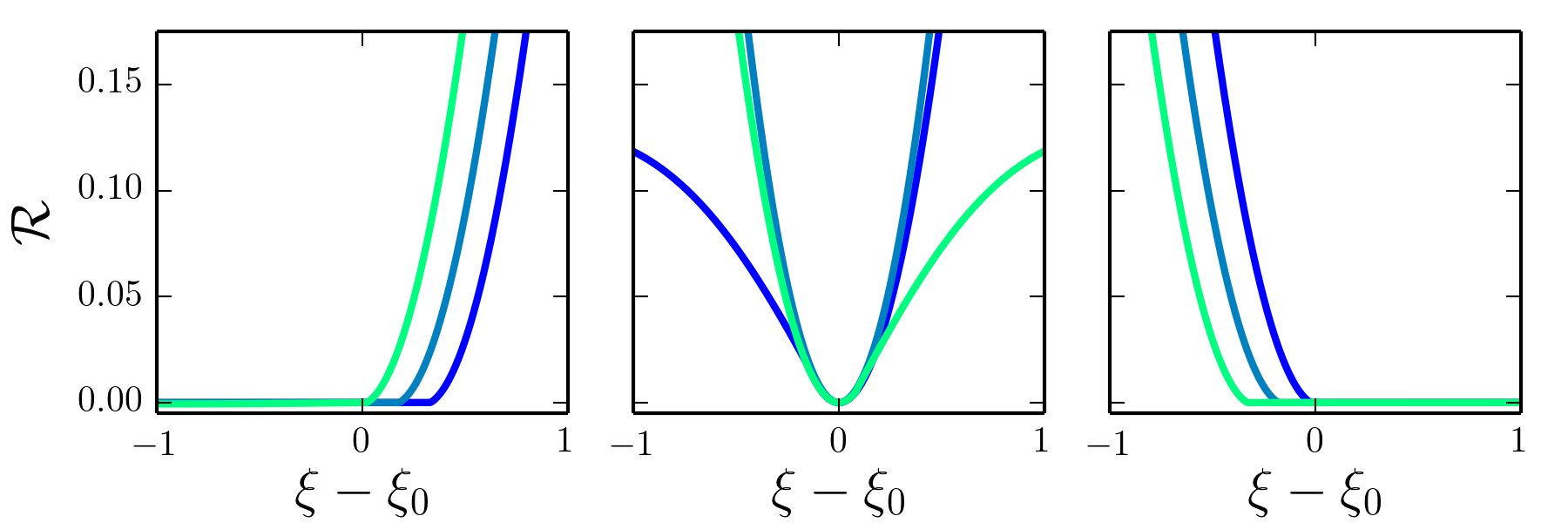}

   \caption{The comoving curvature perturbation $\mathcal{R} (\xi)$ for collisions with $\Delta \xsep =1$ (top) and $\Delta \xsep =2$ (bottom) for observation-side instanton-born observers (left), overlap-born observers (center), and collision-side instanton-born observers (right). Curves from blue to green are for increasing values of reference position $x_0$. For collisions between identical bubbles, $\xi_0$ need not increase monotonically with $x_0$. More specifically, on the top in the left panel we sample $x_0 = -0.90, -0.75, - 0.60$ corresponding to $\xi_0=-1.05, -0.83, -0.64$, in the centre panel we sample $x_0 = -0.5, 0.5, 1.5$ corresponding to $\xi_0 = -0.56, 0.00, 0.56$, and in the right panel we sample $x_0 = 1.60, 1.75, 1.90$ corresponding to $\xi_0 = 0.63, 0.83, 1.00$. On the bottom in the left panel we sample $x_0 = 0.10, 0.25, 0.40$ corresponding to $\xi_0=0.10, 0.25, 0.40$, in the centre panel we sample $x_0 = 0.5,1.0, 1.5$ corresponding to $\xi_0 = 0.33,0.00,-0.34$, and in the right panel we sample $x_0 = 1.60, 1.75, 1.90$ corresponding to $\xi_0 = -0.41, -0.25, -0.10$.
   }
   \label{fig:identical_observers}
\end{figure}

The advantage of the method used in this paper is our ability to explore cosmological observables from any vantage point inside the bubble. In Fig.~\ref{fig:identical_observers} we show the comoving curvature perturbation in the vicinity of observation-side instanton-born observers (left panel), observation-side and collision-side overlap-born observers (center panel), and collision-side instanton-born observers, for $\Delta \xsep =1$ (top) and $\Delta \xsep =2$ (bottom). Instanton-born observers see a translated version of the comoving curvature perturbation. A subset of these observers will have causal access to the collision at late times, and predictions for the comoving curvature perturbation in these cases matches previous work~\cite{Wainwright:2013lea,Wainwright:2014pta}, as explicitly shown in Fig.~\ref{fig:geo_comparison}. 

Overlap-born observers experience a non-zero, approximately planar symmetric, curvature perturbation everywhere. The shape of this perturbation is shown in the center panel of Fig.~\ref{fig:identical_observers}. A subset of these observers, the near-boundary overlap-born observers, have causal access to the collision boundary. From Fig.~\ref{fig:identical_observers}, these near-boundary overlap-born observers see a nearly linear comoving curvature perturbation on one side of the collision boundary, and a nearly quadratic comoving curvature perturbation on the other side of the collision boundary, spreading the signal out over the whole CMB sky. This is in contrast to the near-boundary instanton-born observers, who would record a zero curvature perturbation on one side of the collision boundary, and a rising curvature perturbation on the other, as in the left panel of Fig.~\ref{fig:identical_observers}. This gives rise to a localized signal on the CMB sky.

This result addresses several conjectures made in previous work \cite{Aguirre:2008wy,Aguirre:2009ug,Chang_Kleban_Levi:2009,Czech:2010rg,Gobbetti_Kleban:2012,Kleban_Levi_Sigurdson:2011,Kleban:2011pg,Feeney_etal:2010dd,Feeney_etal:2010jj}, which argued that 1) the collision induces a very nearly linear curvature perturbation that turns on at the collision boundary and consequently 2) that near-boundary overlap-born observers\footnote{In previous papers these were named ``foreign-born", while instaton-born observers were ``native-born."} would experience a comoving curvature perturbation that was a mirror image of the perturbation experienced by instanton-born observers on the other side of the collision boundary.~\footnote{Let us briefly explain how the second conjecture follows from the first. A purely linear curvature perturbation in an otherwise homogeneous Universe can be gauged away~\cite{Hinterbichler:2012nm}. However, in the collision spacetime, where a linear curvature perturbation is matched to a region with no curvature perturbation across the collision boundary, the curvature perturbation cannot be gauged away. Rather, the linear curvature perturbation in the collision region can be gauged away at the expense of inducing a linear perturbation in the region not affected by the collision. Observers an equal distance from the collision boundary on either side would therefore see an identical perturbation.} The first conjecture was addressed in the second paper of this series~\cite{Wainwright:2014pta}, where we showed that a previously neglected contribution from slow-roll inflation inside each bubble gives rise to a curvature perturbation that is quadratic in the distance from the collision boundary. Because a quadratic curvature perturbation cannot be gauged away, the signal from a collision observed by overlap-born observers is not localized on the sky. Therefore, instanton-born and overlap-born near-boundary observers do not record mirror-image signals.

Deep into the collision region, for the far-from-boundary overlap-born observers, only the quadratic part of the perturbation is causally accessible. The leading observable for far-from-boundary overlap-born observers is therefore a CMB quadrupole, as described in Sec.~\ref{sec:CMB_observables}. Comparing the outcome for the two initial separations, we see that a larger $\Delta \xsep$ yields steeper profiles for $\mathcal{R} (\xi)$. This is in agreement with previous work~\cite{Wainwright:2014pta}, where increasing $\Delta \xsep$ was found to lead to an increasing slope of the comoving curvature perturbation as seen by instanton-born observers. Here, we show that the same is true for overlap-born observers.

\begin{figure}
   \centering
   \includegraphics{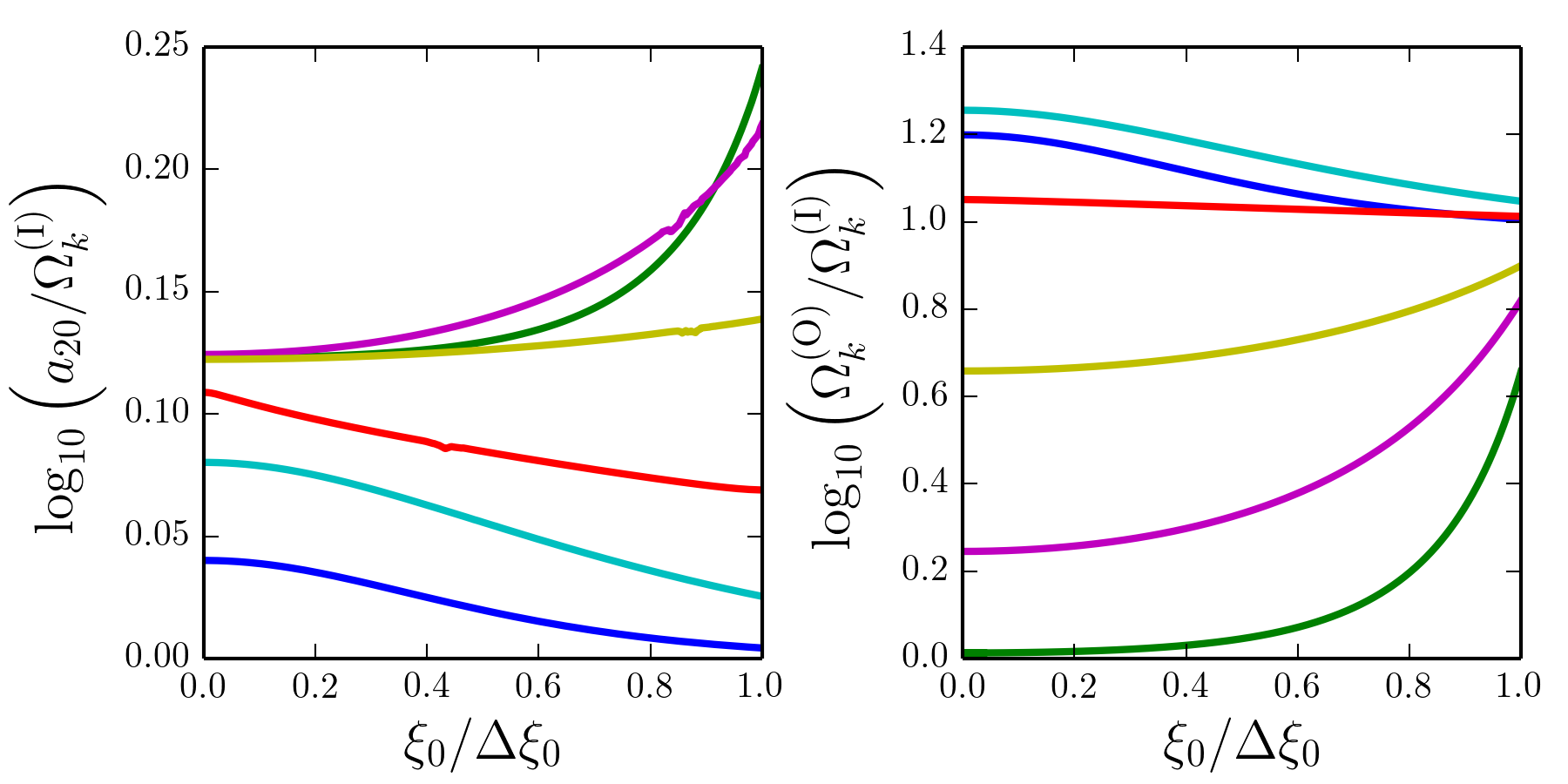}
   \caption{Predictions for $a_{20}$ (left) and $\Omega_k^{\rm (O)}$ (right) seen by overlap-born observers in collisions between identical bubbles. Curves correspond to initial separations $\Delta \xsep = 0.5, 1.0, 1.5, 2.0, 2.5, 3.0$ (blue, cyan, red, yellow, purple, green). We plot observables as a function of the fractional distance in positions $\xi_0$ from the centre of the collision (the left hand side of each panel) to the edge of the overlap-born region (the right hand side of each panel).
}
   \label{fig:identical_observers_a20_omegak}
\end{figure}

In Fig.~\ref{fig:identical_observers_a20_omegak} we show the $\Omega_k^{\rm (O)}$ and $a_{20}$ measured by overlap-born observers for collisions with kinematics $\Delta \xsep = 0.5, 1.0, 1.5, 2.0, 2.5, 3.0$. The overlap-born region in each case has finite extent (in proper distance). Different kinematics yield overlap-born regions of different size. In addition, observables are symmetric about the position of the collision. For ease of comparison, we therefore plot $\Omega_k^{\rm (O)}$ and $a_{20}$ as a function of the fractional distance in observer positions $\xi_0$ from the centre of the collision (the left hand side of each plot in Fig.~\ref{fig:identical_observers_a20_omegak}) to the edge of the overlap-born region (the right hand side of each plot in Fig.~\ref{fig:identical_observers_a20_omegak}).

For small $\Delta \xsep$, the negative spatial curvature is larger than in the instanton-born region, while for large $\Delta \xsep$ the curvature is smaller. Comparing the contours of constant field for increasing $\Delta \xsep$ in Fig.~\ref{fig:identical_observers_contours}, we see that they are convex in the overlap-born region for $\Delta \xsep =1$ and concave for $\Delta \xsep =2$. In the cases where the negative curvature is higher than the instanton-born region, $a_{20}$ is maximized at the center of the overlap-born region. When the curvature is lower, $a_{20}$ is minimized at the center of the overlap-born region. In general, increasing initial separation yields a larger magnitude for $a_{20}$. Note also that the prediction is for $a_{20}$ to be positive definite for identical bubble collisions.

A particular model of the scalar potential will yield an ensemble of correlated quadrupoles and curvatures, corresponding to collisions with varying $\Delta \xsep$ and varying observer position. Importantly, not every region in this parameter space is covered by a particular model. With further assumptions, one can put a prior on this parameter space. In Refs.~\cite{Wainwright:2013lea}, the prior over bubble separations was shown to be ${\rm Pr} (\Delta \xsep) \propto \sin^3 \Delta \xsep$. Roughly $\sim 95 \%$ of the prior falls between the $0.5 < \Delta \xsep < 2.5$ curves in Fig.~\ref{fig:identical_observers_a20_omegak}.

\begin{figure}
   \centering
   \includegraphics{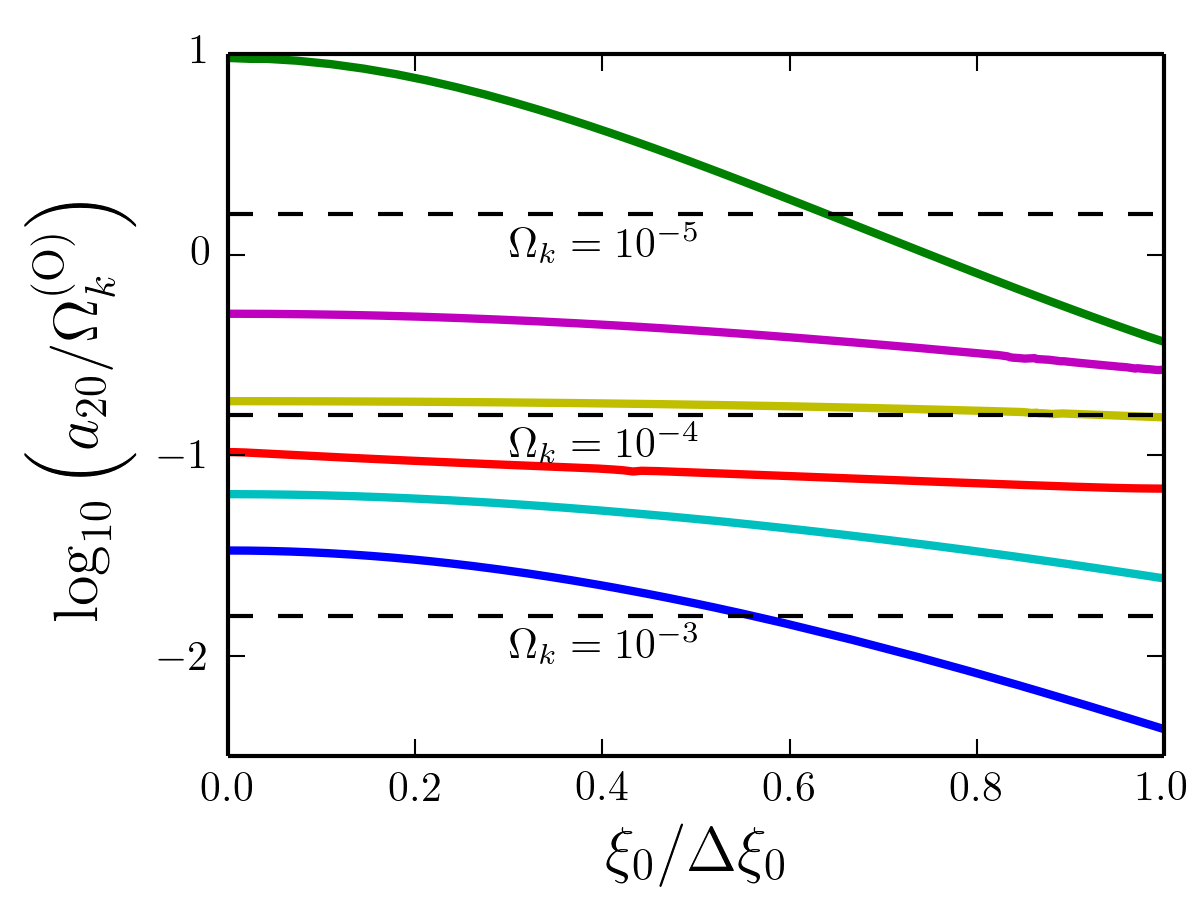}
   \caption{Predictions for the ratio of $a_{20}$ and $\Omega_k^{\rm (O)}$ seen by overlap-born observers in collisions between identical bubbles. Curves correspond to initial separations $\Delta \xsep = 0.5, 1.0, 1.5, 2.0, 2.5, 3.0$ (blue, cyan, red, yellow, purple, green). The predicted values of this ratio using the observed magnitude of the CMB quadrupole and hypothetical measurements of curvature in the range $10^{-3} < \Omega_k < 10^{-5}$ are shown as dashed horizontal lines.}
   \label{fig:identical_observers_ratio}
\end{figure}

Let us briefly comment on the ability of overlap-born observers to constrain scalar field models using observations of curvature and the CMB quadrupole. First, one can alter a scalar field potential to add more $e$-folds of inflation, which scales down $\Omega_k^{\rm (I)}$ to any desired level. This can make a model that is compatible with a non-observation of curvature by overlap-born observers.
However, should negative curvature be observed, then one can make useful statements about the hypothesis that we could be an overlap-born observer. It is useful to use the ratio of the CMB temperature quadrupole and the spatial curvature as a proxy, since this is independent of the unspecified instanton-born curvature parameter $\Omega_k^{\rm (I)}$. We show the predictions for this ratio in Fig.~\ref{fig:identical_observers_ratio} using our fiducial model. Assuming that the contribution to the CMB quadrupole cannot be larger than the observed value $a_{20} \simeq 1.6 \times 10^{-5}$,\footnote{One possible explanation for the anomalously low observed CMB quadrupole is interference between a contribution from pre-inflationary initial conditions and the subsequent contribution from fluctuations during inflation. Here, we are assuming that the two contributions are of the same order of magnitude, and do not have any finely tuned cancelations.} we can compare with different scenarios for an observed negative curvatures ranging from $10^{-3} < \Omega_k < 10^{-5}$ (dashed lines in Fig.~\ref{fig:identical_observers_ratio}).  If there is an intersection between the predicted curves and the observed ratio, a particular model would be viable, and one would obtain an estimate of the range of compatible $\Delta \xsep$ in that model. In particular, the example shown in Fig.~\ref{fig:identical_observers_ratio} would be viable for any observed curvature in this range. Predictions from other models may have no intersection with the observed ratio, in which case one could rule out the possibility that we are an overlap-born observer in such a model.

\section{Simulating collisions between non-identical bubbles}

We now discuss collisions between non-identical bubbles. In this case, the colliding bubbles contain different vacua, and consequently, bubbles do not merge; instead a domain wall that separates the bubble interiors must form after the collision. The dynamics of the post-collision domain wall play an important role in determining the structure of the post-collision spacetime. These dynamics are straightforward to assess in the thin-wall limit for vacuum bubbles: for an observation bubble with Hubble parameter $H_o$, a collision bubble with Hubble parameter $H_C$ and an inter-bubble domain wall with surface tension $\sigma_{Co}$, we have the following relation:
\begin{eqnarray}
H_C^2 - H_o^2 + 16 \pi^2 \sigma_{Co}^2 &>& 0, \ \ \ {\rm Wall \ accelerates \ into \ collision \ bubble} \nonumber \\
H_C^2 - H_o^2 + 16 \pi^2 \sigma_{Co}^2 &<& 0, \ \ \ {\rm Wall \ accelerates \ into \ observation \ bubble} \nonumber
\end{eqnarray}
When the surface tension is subdominant to the energy splitting, the conclusion is that the bubble with a lower vacuum energy expands into the bubble with a higher vacuum energy. In the following, we do not consider vacuum bubbles, but rather consider bubbles with an inflationary interior (as in the previous section). However, the qualitative results of the thin-wall analysis hold since the Hubble parameter remains roughly constant during inflation. 

Numerical analyses of the case in which the wall accelerates into the collision bubble have been performed both with~\cite{Johnson:2011wt,Wainwright:2014pta} and without~\cite{Aguirre:2008wy,Chang_Kleban_Levi:2009} full GR in previous literature. In Ref.~\cite{Wainwright:2014pta} the signature for instanton-born observers was determined. However, there are a number of important questions left open regarding the overlap-born observers. In particular, previous work~\cite{Aguirre:2008wy,Aguirre:2009ug,Chang_Kleban_Levi:2009,Czech:2010rg,Gobbetti_Kleban:2012,Kleban_Levi_Sigurdson:2011,Kleban:2011pg,Feeney_etal:2010dd,Feeney_etal:2010jj} speculated that overlap-born observers with causal access to the collision boundary would have identical observables to their neighbouring instanton-born cousins. In the last section, we have shown this to be false for collisions between identical bubbles, and below we show it to be false for non-identical bubble collisions as well. 

Another assertion made in previous work~\cite{Aguirre:2008wy,Aguirre:2009ug} was that the reheating surface in the overlap-born region would be everywhere spacelike, and infinite in spatial extent. It was further argued that far-from-boundary overlap-born observers would experience a very nearly homogeneous and isotropic universe.  These assertions were made regarding the bubble away from which the domain wall accelerates, and we find below using our simulations that there is good evidence that these speculations are indeed true. 

In all previous work on bubble collisions, the case where the wall accelerates into the observation bubble was considered fatal, and thought to allow no viable observation-side far-from-boundary overlap observers. Below, we show this assumption to be invalid, and that the results are qualitatively similar to the case where the wall accelerates into the collision bubble. The late-time surfaces of constant field during inflation inside the observation bubble simply re-adjust so as to become everywhere spacelike.

\subsection{Domain wall accelerates into the collision bubble}

\begin{figure}
   \centering
   \includegraphics{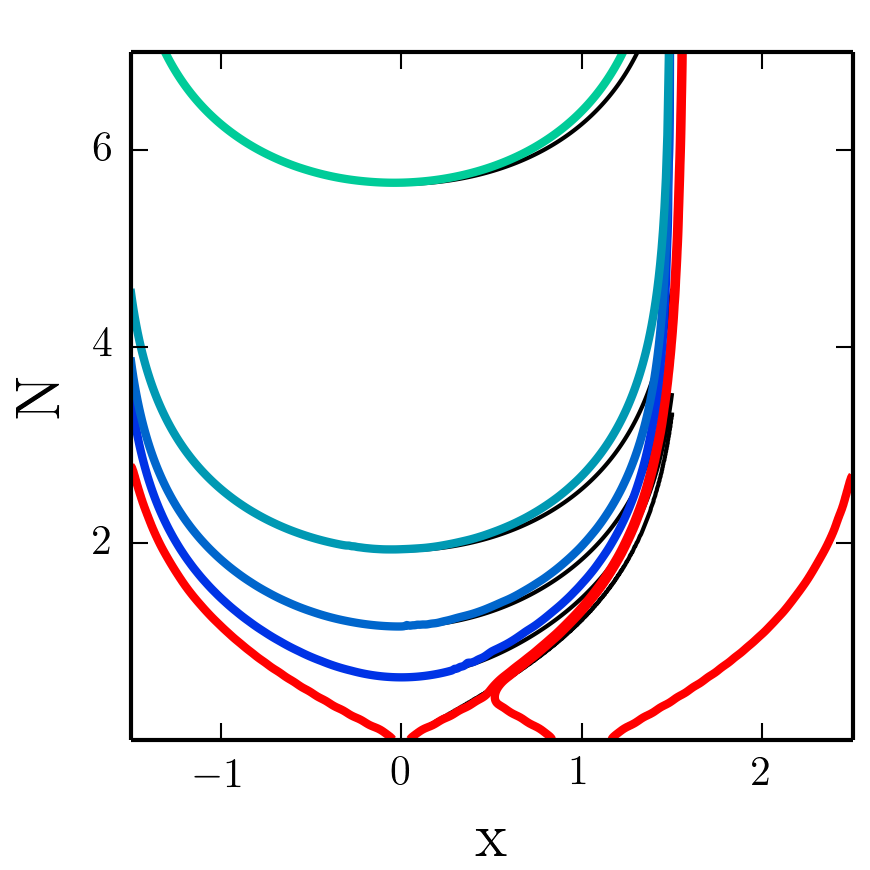}
   \caption{A contour plot of $\phi(x, N)$ for the collision between non-identical bubbles with an initial separation $\Delta \xsep =1$. The red contours ($\phi = \pm 10^{-4}\, \Mpl$)  tracks the bubble walls, successive hypersurfaces ($\phi = 0.005, 0.013, 0.03, 0.13\, \Mpl$) are blue to green contours, and the corresponding contours in an unperturbed spacetime are shown in grey. }
   \label{fig:nonidentical_contourplot}
\end{figure}

We first consider the case where the domain wall accelerates into the collision bubble. We perform a simulation using the same quartic barrier potential as in Sec.~\ref{sec:identical}, but with initial conditions containing the two different types of bubbles that can be nucleated from the false vacuum. The observation bubble contains the same inflationary plateau as in the previous section, while the collision bubble is a vacuum bubble with (large) positive cosmological constant. In this case, a domain wall separating the interior of the two bubbles forms after the collision. The domain wall quickly accelerates into the collision bubble, as can be seen in the contour plot Fig.~\ref{fig:nonidentical_contourplot}. Here, we show the same contours as Fig.~\ref{fig:identical_observers_contours}. We also over-plot contours for an unperturbed observation bubble. 

Comparing with Fig.~\ref{fig:identical_observers_contours}, there are two important differences of note between identical and non-identical bubble collisions. First, the difference between the perturbed and unperturbed observation bubble is far smaller for the non-identical bubble collision shown here. This is an example of a ``mild" collision in the parlance of Ref.~\cite{Aguirre:2007wm}, and is expected to have a minimal effect on cosmological observables (which is indeed shown to be true below). Second, the surfaces of constant field are advanced with respect to those in the unperturbed bubble, while they were retarded in the identical bubble collisions studied above. This implies that the sign of the comoving curvature perturbation will be opposite to that found for identical bubble collisions, an observation made previously in Refs.~\cite{Chang_Kleban_Levi:2009,Johnson:2011wt,Gobbetti_Kleban:2012,Wainwright:2013lea,Wainwright:2014pta}.

\begin{figure}
   \centering
   \includegraphics{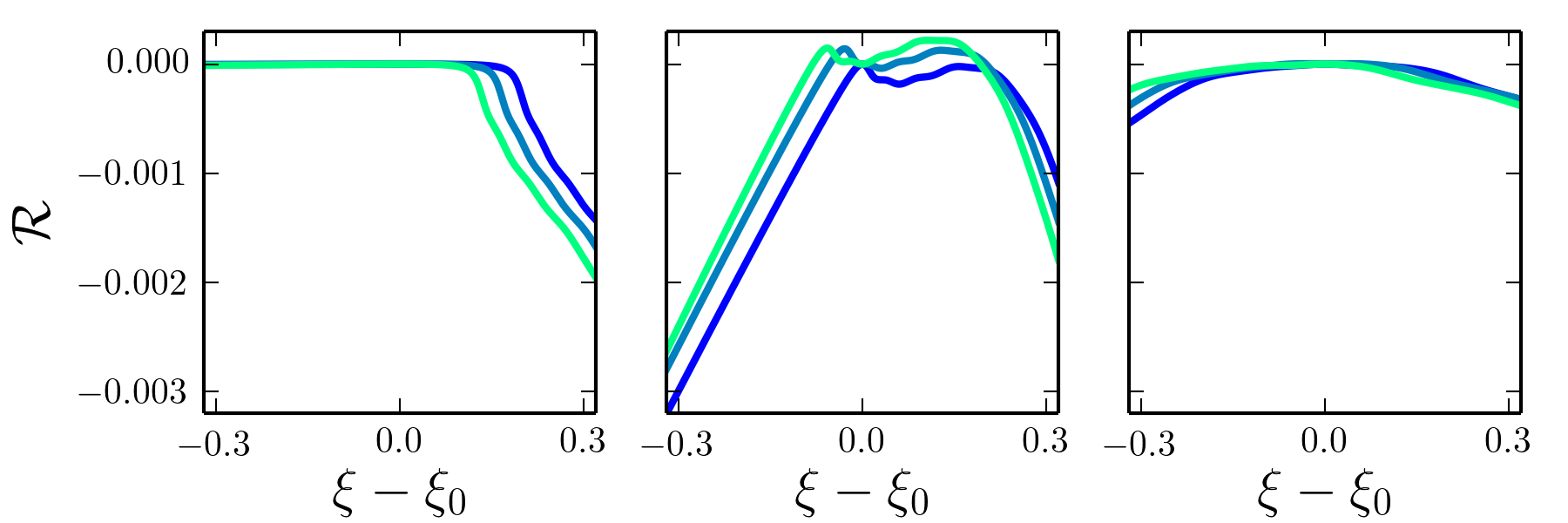}
   \caption{The comoving curvature perturbation $\mathcal{R} (\xi)$ for collisions with $\Delta \xsep =1$ between non-identical bubbles with a post-collision domain wall that accelerates into the collision bubble. Panels correspond to near-boundary instanton-born observers (left), near-boundary overlap-born observers (center), and far-from-boundary overlap-born observers  (right). Curves from blue to green are for increasing values of reference position $x_0$.  There is a one-to-one relation between $x_0$ and $\xi_0$ for non-identical bubble collisions. In the left panel we sample $x_0 = -0.725, -0.700, -0.675$ corresponding to $\xi_0=-0.79, -0.76,-0.73$, in the centre panel we sample $x_0 = -0.575,-0.550, -0.525$ corresponding to $\xi_0 = -0.59,-0.57,-0.54$, and in the right panel we sample $x_0 = 1.000, 1.025, 1.050$ corresponding to $\xi_0 = 1.38, 1.44, 1.50$.
   }
   \label{fig:nonidentical_observers}
\end{figure}

Extracting the comoving curvature perturbation on the $\phi_0 = 0.3\, \Mpl$ hypersurface at different positions, in Fig.~\ref{fig:nonidentical_observers} we sample near-boundary instanton- (left panel) and overlap- (center panel) born observers, as well as far-from-boundary  overlap-born ones (right panel). Note that these are all observation-side observers. As expected, instanton-born observers at different positions see a translated version of the comoving curvature perturbation. In this example, the comoving curvature perturbation has visible oscillations due to internal breathing modes of the post-collision domain wall that are excited by the collision~\cite{Wainwright:2014pta}. Comparing with the curvature perturbation obtained for identical bubble collisions in Fig.~\ref{fig:identical_observers}, we see that here it takes the opposite sign and is nearly two orders of magnitude smaller in amplitude. Near-boundary overlap-born observers see a nearly linear comoving curvature perturbation across the collision boundary, and a non-zero, growing perturbation in the opposite direction. Note that the structure seen in the comoving curvature perturbation due to the internal wall modes is visible here as well. Moving deeper into the overlap region, local FRW patches become increasingly homogeneous. Note that the oscillations are present in this example fairly deep into the collision region, implying that the internal wall modes take some time to ring down.

\begin{figure}[t]
   \centering
   \includegraphics{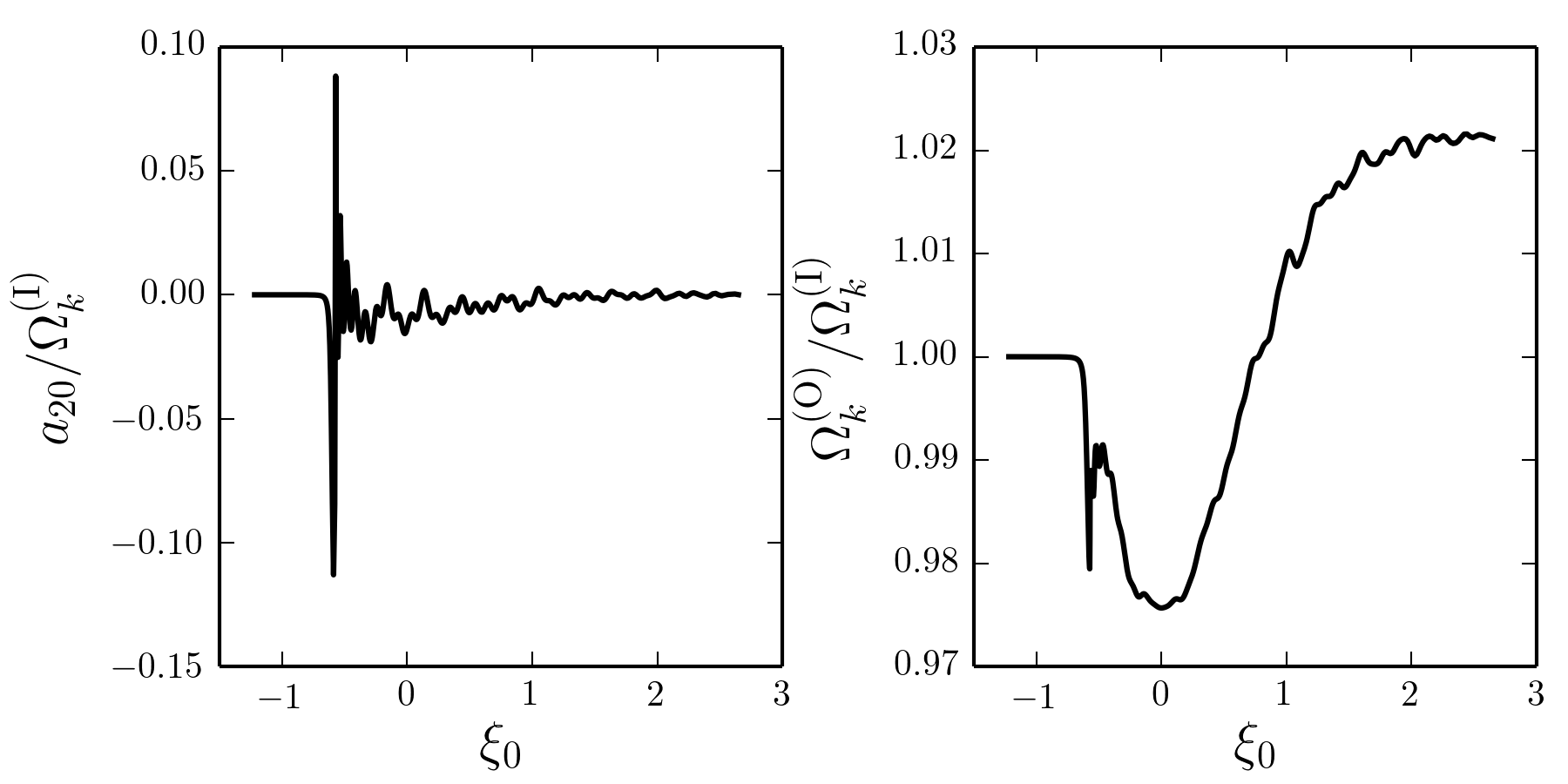}
   \caption{The CMB quadrupole (left) and spatial curvature (right) experienced by observers at various positions in the collision spacetime. Deep into the collision region, locally the universe becomes increasingly homogeneous with a constant value of the curvature.}
   \label{fig:nonidentical_a2_omega}
\end{figure}

We now turn to an exploration deep into the collision region. Tracking the $\phi_0=0.3\, \Mpl$ hypersurface deep into the collision region, we find that it is everywhere spacelike in the simulation. To determine if there are indications that the slice may become timelike at some point outside the simulation, we find the quadrupole and spatial curvature at all reference points on the $\phi_0=0.3\, \Mpl$ slice. This is shown in Fig.~\ref{fig:nonidentical_a2_omega}. As one moves deeper into the overlap-born region, the local FRW patch becomes increasingly homogeneous, as evidenced by the decreasing magnitude of the observed quadrupole. The spatial curvature first decreases slightly from its value in the instanton-born region, and then increases. The variation in curvature is only at the percent level, and the curvature appears to asymptote to a constant value deep into the collision region. If the local FRW patches had become increasingly inhomogeneous, that would have been an indication that the constant-field hypersurfaces might become timelike at some point outside the simulation. Extrapolating the results of our simulation, it therefore appears that there is no obstruction to the far-from-boundary overlap-born region being infinite in spatial extent. 

\begin{figure}
   \centering
   \includegraphics{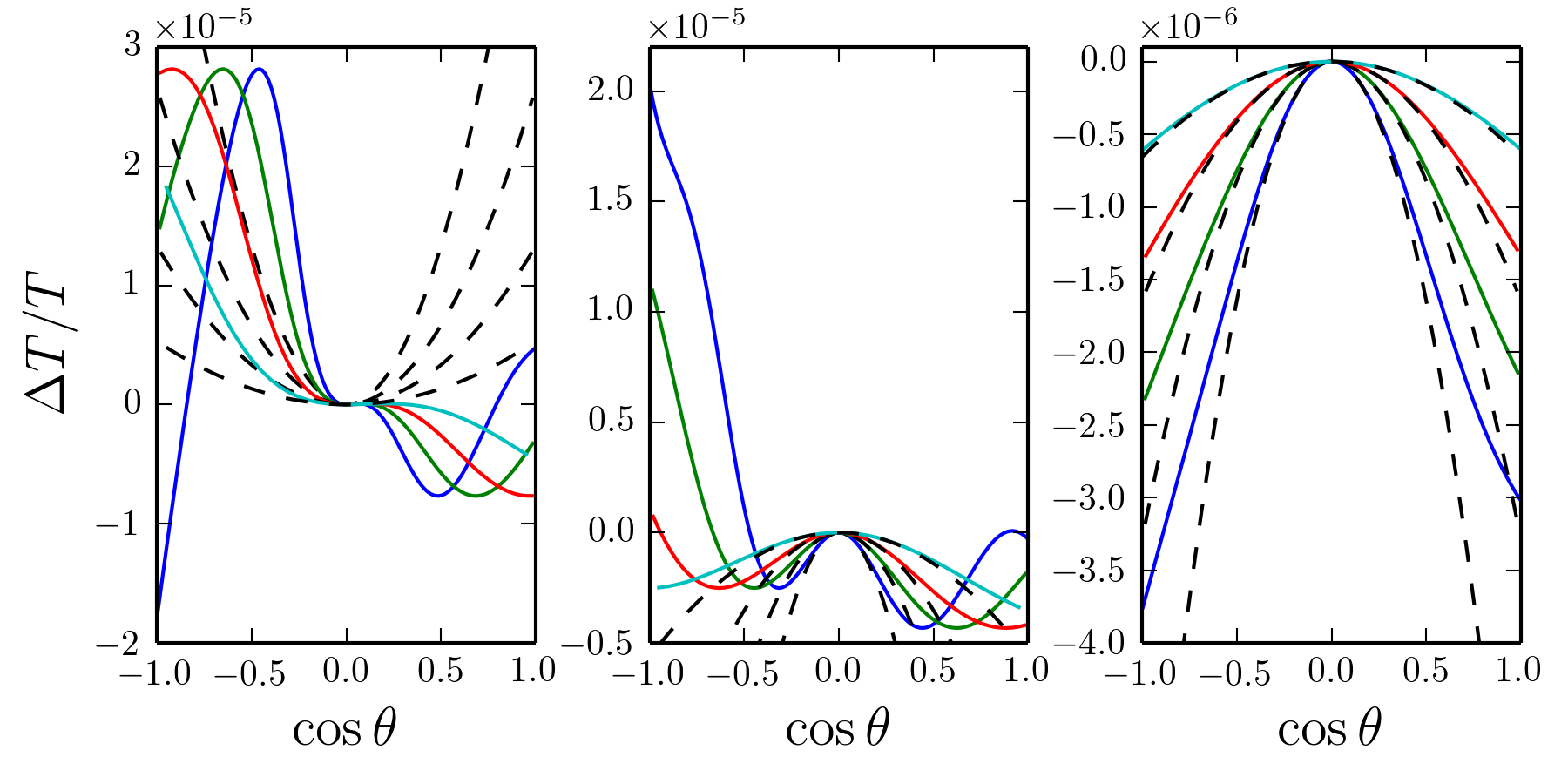}
   \caption{The temperature anisotropies seen by overlap-born observers with observed curvature $\Omega_K^{\rm (I)} = 0.001, 0.0005, 0.00025, 0.0001$ (blue, green, red, cyan) at three positions: near the collision boundary $x_0=-0.55$ corresponding to $\xi_0 = -0.20$ (left panel), far from the collision boundary $x_0=1.0$ corresponding to $\xi_0=1.5$ (right panel), and at an intermediate position $x_0= 0.5$ corresponding to $\xi_0 = 0.5$ (centre panel). The locally computed temperature quadrupole (Eq.~\ref{eq:multipoles}) is shown for each curvature and observer position as dashed lines.}
   \label{fig:nonidentical_multipoles}
\end{figure}

The rapid variation of the locally observed quadrupole in Fig.~\ref{fig:nonidentical_a2_omega} indicates that there is information in higher multipoles as long as $\Omega_K^{\rm (I)}$ is not too small. In Fig.~\ref{fig:nonidentical_multipoles}, we show the full projected temperature anisotropy in the Sachs-Wolfe approximation (Eq.~\ref{eq:dtsachswolfe}) seen by overlap-born observers at three different positions, for $\Omega_K^{\rm (I)} = 0.001, 0.0005, 0.00025, 0.0001$ (below current constraints, but possibly observable). In each, we overplot the temperature anisotropy given by the quadrupole moment as a dashed line. For observers near the collision boundary, the wall modes leave a significant amount of visible structure in the temperature anisotropies over this range of curvatures. It is only deep into the collision region, or for smaller curvatures, that the quadrupole is a good characterization of the temperature anisotropies. Overlap-born observers near enough to the collision boundary, and with large enough spatial curvature, would see an observationally significant large angular scale contribution to the temperature anisotropies from wall modes. Decomposing into multipoles, the planar symmetry of the collision translates into an alignment of the low-$\ell$ spherical harmonic coefficients. It has not escaped the authors' attention that this could be related to various persisting low-$\ell$ anomalies in the CMB such as the so-called ``Axis of Evil"~\cite{Land:2005ad}. However, we defer a more detailed exploration of wall modes to future work.

In contrast to the collision between identical bubbles, where the deviations from the instanton region were nonzero for all overlap-born observers, there is strong evidence that non-identical bubble collisions produce an infinite volume in which overlap-born observers would see no trace of the collision. Only those  in the vicinity of the collision boundary would have the opportunity to access information about the collision in their past.

\subsection{Domain wall accelerates into the observation bubble}

\begin{figure}
   \centering
   \includegraphics{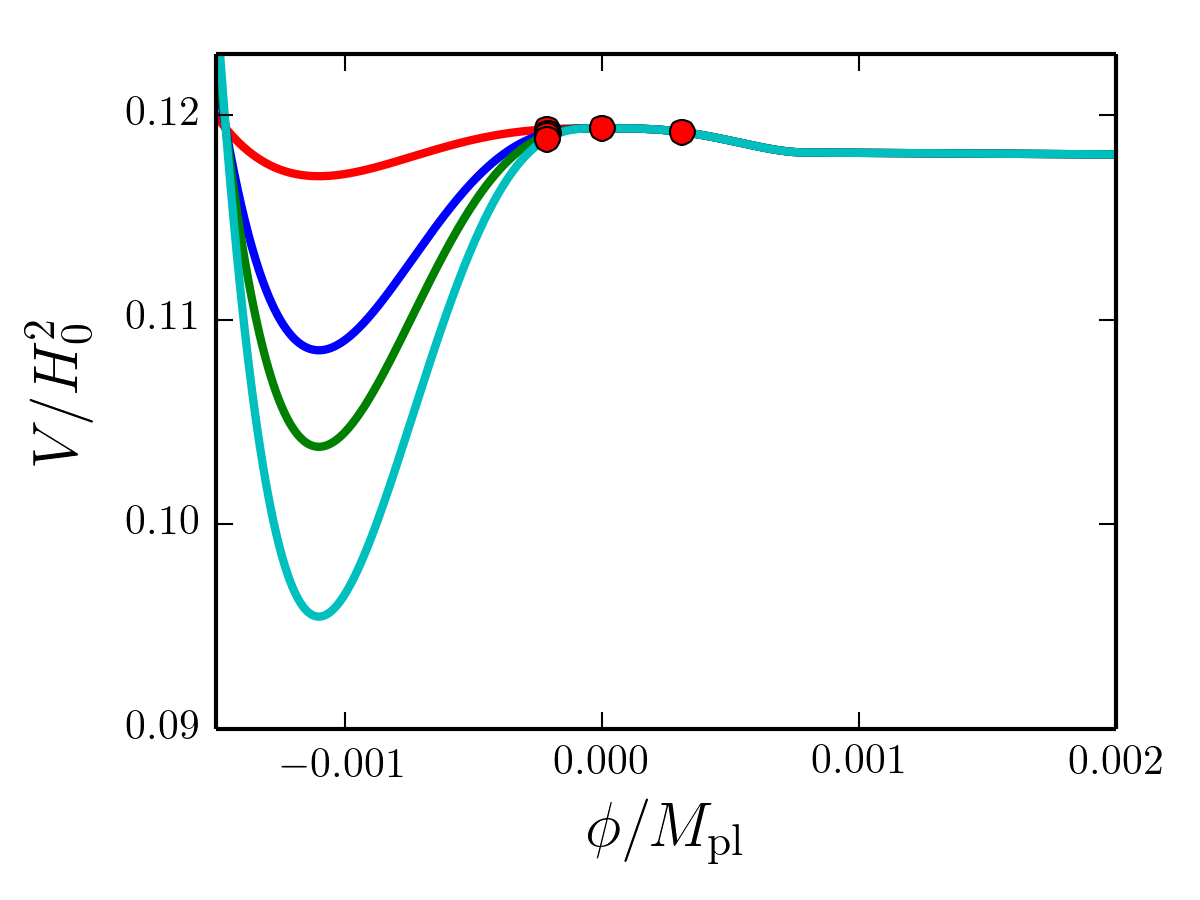}
   \caption{Potentials used in the simulation of non-identical bubble collisions with a post-collision domain wall that accelerates into the observation bubble. The depth of the collision bubble's potential minimum is controlled by the parameter $\mu$, which is set to $\mu = 0.02, 0.1, 0.15, 0.25$ (red, blue, green, cyan). Increasing $\mu$ corresponds to increased depth of the minimum.}
   \label{fig:killer_potentials}
\end{figure}

When the collision bubble contains a lower energy phase, the post-collision domain wall accelerates into the observation bubble. This would appear to be fatal, since the post-collision wall would quickly accelerate to near the speed of light, obliterating everything in its path. However, we saw in the previous section that the surfaces of constant field can re-adjust in the presence of a post-collision domain wall, producing infinite spatial hypersurfaces. To determine what happens, we ran simulations for a set of potentials with varying depth of the collision bubble minimum. The depth of the minimum is controlled by the parameter $\mu$ (see Ref.~\cite{Wainwright:2014pta} for a detailed description of the potential). The potentials, along with the instanton endpoints, are shown in Fig.~\ref{fig:killer_potentials}. 

\begin{figure}[t]
   \centering
   \includegraphics{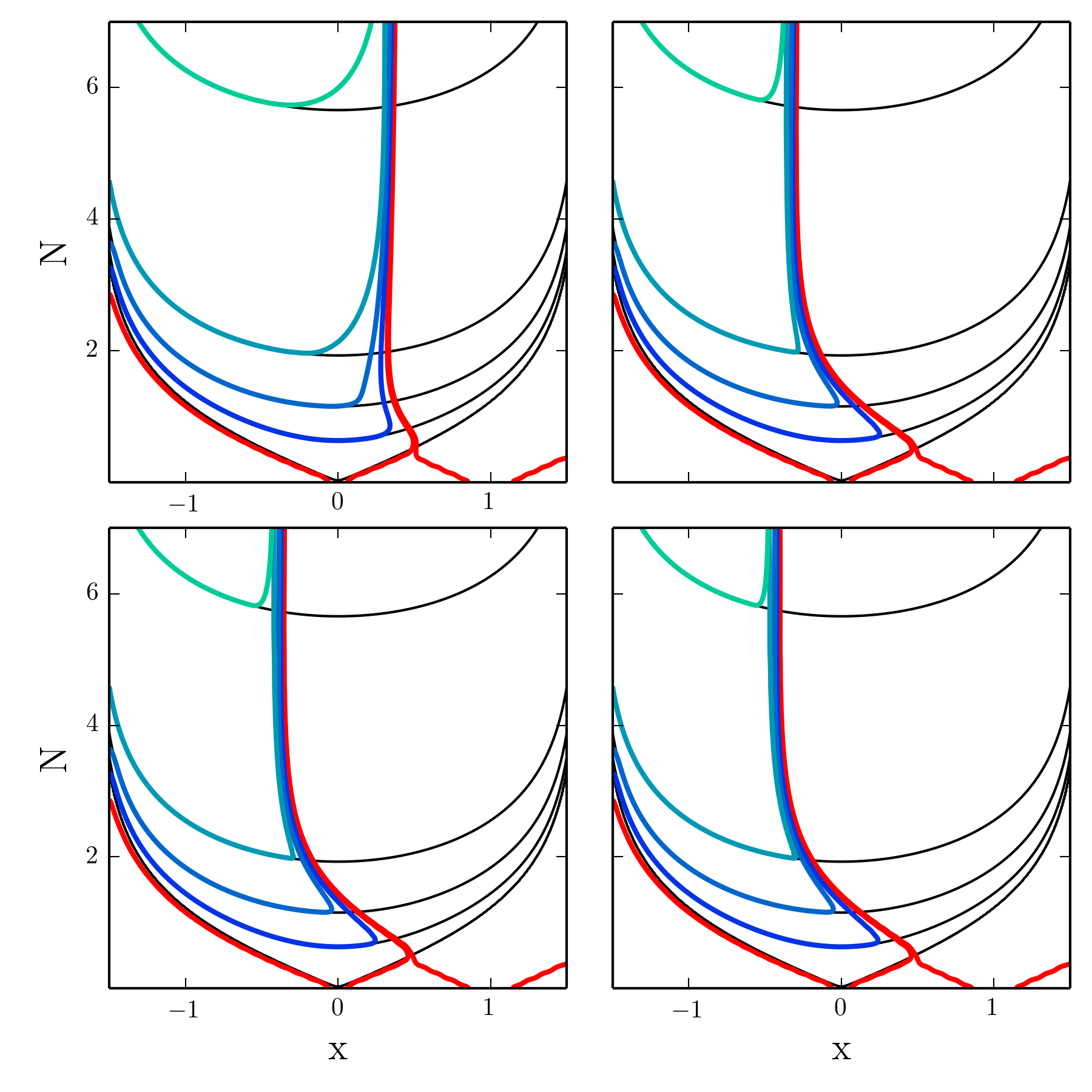}
   \caption{Contour plots of $\phi (N,x)$ for collisions between non-identical bubbles with a post-collision domain wall that accelerates into the observation bubble. The corresponding potentials are shown in Fig.~\ref{fig:killer_potentials}. In each case, the red contour ($\phi = \pm 10^{-4}\, \Mpl$) tracks the bubble wall, successive  hypersurfaces ($\phi = 0.005, 0.013, 0.03, 0.13\, \Mpl$) are blue to green contours, and the corresponding contours in an unperturbed spacetime are shown in grey.}
   \label{fig:killer_contour_plots}
\end{figure}

Contour plots of $\phi (N,x)$ for each of these potentials, with an initial bubble separation of $\Delta \xsep = 1$, are shown in Fig.~\ref{fig:killer_contour_plots}. The degree to which the post-collision domain wall accelerates into the observation bubble increases with the depth of the collision bubble's potential minimum. Early constant-field hypersurfaces clearly are not everywhere spacelike, as they bend back to follow the domain wall. However, at sufficiently large $\phi$, (by $\phi_0 = 0.13\, \Mpl$ in even the most extreme case) the surfaces of constant field do become spacelike. It appears that just as in the case where the post collision domain wall accelerates out of the observation bubble, here too the constant field surfaces re-adjust to become spacelike at sufficiently late times. 

\begin{figure}[t]
   \centering
   \includegraphics{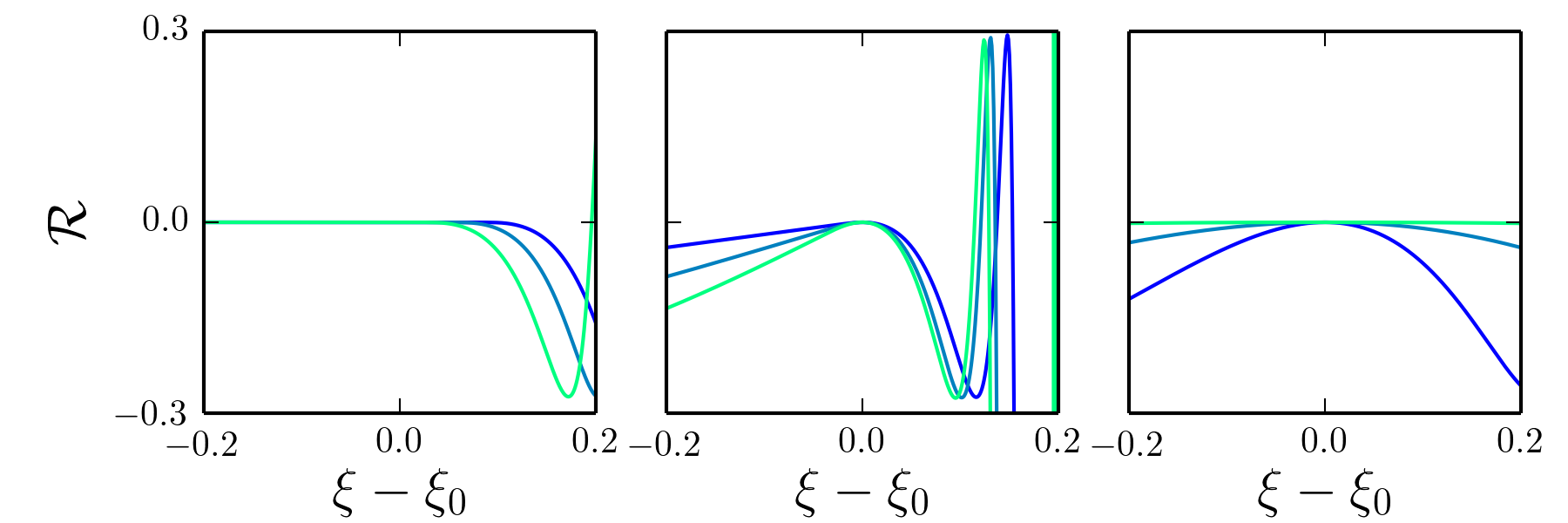}
   \caption{The comoving curvature perturbation $\mathcal{R} (\xi)$ for collisions with $\Delta \xsep =1$ between non-identical bubbles with a post-collision domain wall that accelerates into the observation bubble. Panels correspond to instanton-born observers (left), near-boundary overlap-born observers (center), and far-from-boundary overlap-born observers (right). Curves from blue to green are for increasing values of reference position $x_0$. There is a one-to-one relation between $x_0$ and $\xi_0$ for non-identical bubble collisions. In the left panel we sample $x_0 = -0.650, -0.625, -0.600$ corresponding to $\xi_0=-0.70, -0.67,-0.63$, in the centre panel we sample $x_0 = -0.56,-0.55, -0.54$ corresponding to $\xi_0 = -0.40,-0.14,0.21$, and in the right panel we sample $x_0 = -0.500, -0.460, -0.425$ corresponding to $\xi_0 = 1.9, 3.4, 6.6$.
   }
   \label{fig:killer_observables}
\end{figure}

Taking the particular example of $\mu=0.15$, we show the comoving curvature perturbation at different positions along the $\phi_0 = 0.3\, \Mpl$ hypersurface in Fig.~\ref{fig:killer_observables}. Instanton-born observers (left panel) observe an initially decreasing comoving curvature perturbation. This is in agreement with the fact that the surfaces of constant field in Fig.~\ref{fig:killer_contour_plots} are advanced with respect to the unperturbed observation bubble. Comparing with the identical and non-identical collisions studied above, the magnitude of the comoving curvature perturbation is largest in this case. This makes intuitive sense, as the constant-field hypersurfaces are most dramatically different from the unperturbed bubble here. Passing into the collision region, overlap-born near-boundary observers again see a comoving curvature perturbation that is everywhere non-zero and approximately linear on the other side of the collision boundary. Deep into the collision region, overlap-born far-from-boundary observers see a universe that is increasingly homogeneous with distance from the boundary. 

\begin{figure}
   \centering
   \includegraphics{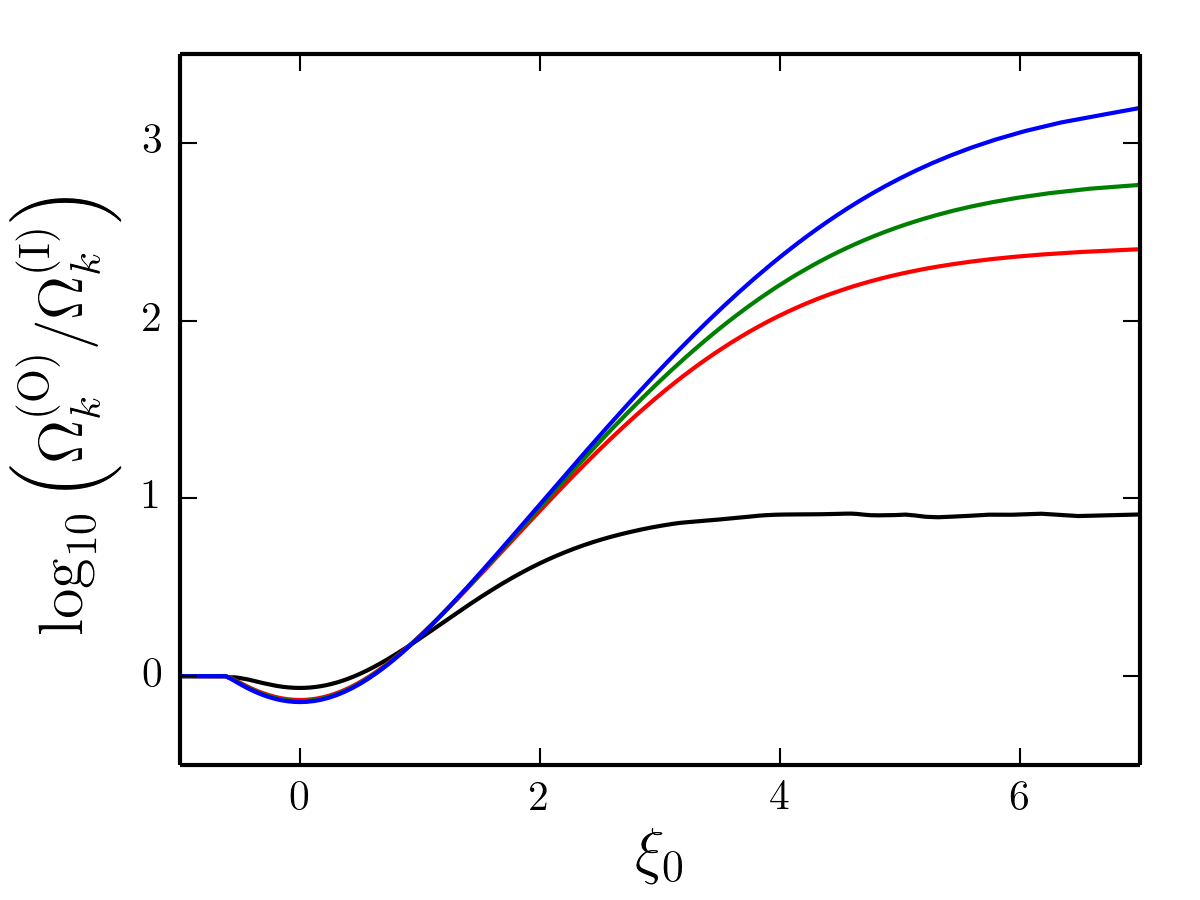}
   \caption{The spatial curvature as a function of $\xi$ in collision spacetimes where the domain wall accelerates into the observation bubble. Curves correspond to potential parameters $\mu = 0.02, 0.1, 0.15, 0.25$ (black, red, green, blue), or equivalently, increasing depth of the collision bubble potential minimum.}
   \label{fig:killer_omegak}
\end{figure}

The spatial curvature asymptotes to a constant for overlap-born far-from-boundary observers, as shown in Fig.~\ref{fig:killer_omegak}. Comparing with the percent-level change in curvature produced by a post-collision domain wall that accelerates into the collision bubble, the result here is dramatic: the curvature can increase for large values of the potential parameter $\mu$ by a factor of $10^3$! From this, we can conclude that the acceleration of the post-collision domain wall into the observation is detrimental to inflation, causing a larger observed curvature. For a sufficiently large acceleration, we can speculate that inflation would be largely disrupted for overlap-born observers, leading to a cosmology that would not be viable as a description of our universe. A study of such models is unfortunately beyond the reach of our numerics due to the large spatial resolution necessary to accurately track highly Lorentz-contracted domain walls associated with large accelerations.

\begin{figure}
   \centering
   \includegraphics{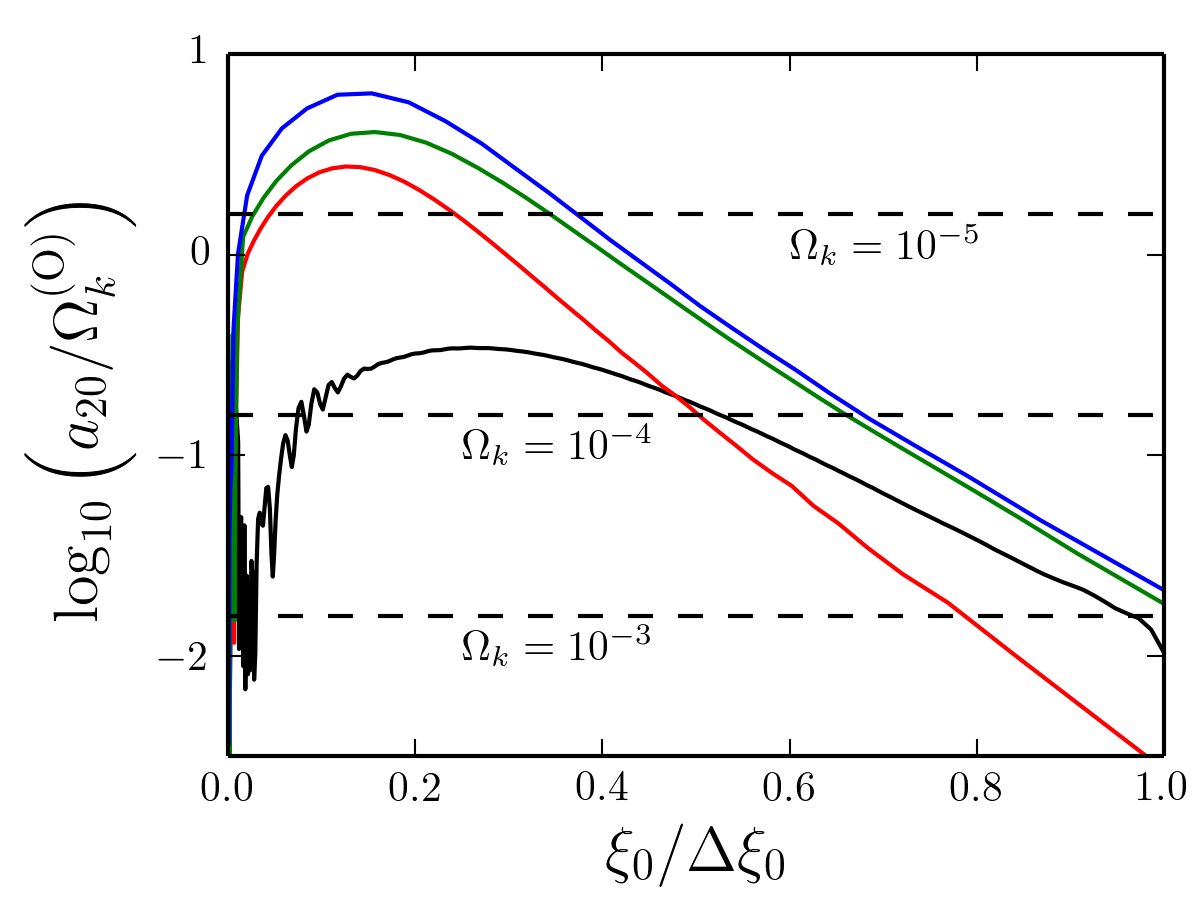} 
   \caption{Predictions for the ratio of $a_{20}$ and $\Omega_k^{\rm (F)}$ seen by overlap-born observers in collisions between non-identical bubbles where the post-collision domain wall accelerates into the observation bubble. Curves correspond to potential parameters $\mu = 0.02, 0.1, 0.15, 0.25$ (black, red, green, blue), or equivalently, increasing depth of the collision bubble potential minimum. The predicted values of this ratio using the observed magnitude of the CMB quadrupole and hypothetical measurements of curvature in the range $10^{-3} < \Omega_k < 10^{-5}$ are shown as dashed horizontal lines.}
   \label{fig:killer_omegak_a2_ratio}
\end{figure}

As described in Sec.~\ref{sec:identical}, models can be tested with the detection of negative spatial curvature by comparing the observed and predicted values for the ratio of the CMB quadrupole and the curvature parameter. The result for the models studied in this section is shown in Fig.~\ref{fig:killer_omegak_a2_ratio}. For scenarios where curvature is detected at the level of $10^{-3} < \Omega_k < 10^{-5}$, essentially all models could accommodate the observation. Recall that there were also collisions between identical bubbles that could accommodate a possible detection of curvature at this level. Although the variation in the magnitude of the quadrupole and curvature are quite different in these two models, the ratio lies in the same range of magnitudes. There is also an important question of distinguishability for overlap-born observers: many models give identical predictions for observables. If the curvature is on the lower end of the allowed values, it may be possible to measure higher multipoles associated with wall breathing modes as described in the previous section.  

Moving into the collision region, the overlap-born far-from-boundary observers experience an increasingly homogeneous universe. Extrapolating the results of the simulation, we therefore conclude that the universe to the future of the collision is infinite in spatial extent. This is a dramatic departure from the expectation that there should be no overlap-born observers! Such collisions are not fatal. In fact, there is seemingly an infinite set of overlap-born observers who live in a nearly homogeneous universe, blissfully ignorant of the dramatic bubble collision in their past. 

\section{Implications for probabilities and measures in eternal inflation}

In an eternally inflating Universe, any given bubble will experience an infinite number of collisions~\cite{Garriga:2006hw}; any specific observer inhabiting a bubble has access to a subset of these collisions due to the finite extent of their particle horizon. The expected number of collisions in the causal past of an observer has been assessed in previous work~\cite{Garriga:2006hw,Aguirre:2007an,Freivogel_etal:2009it} under the assumption that the interior of the observation bubble remains undisturbed by bubble collisions. 

An interesting conclusion of this exercise was that although ``false vacuum" eternal inflation has de Sitter space as a background, the cosmological boundary conditions required to (statistically) determine the bubble distribution also define a preferred reference frame detectable in the collision distribution~\cite{Aguirre:2003ck} accessible to a given observer.  Thus even after an ``eternity", there is a ``memory" of initial conditions for eternal inflation.

These studies also revealed that the total number of collisions is formally divergent for observers far ``up the bubble wall", which are infinitely boosted with respect to the preferred frame.  In a measure weighting by volume on spatial hypersurfaces inside the bubble, these constitute essentially all observers~\cite{Aguirre:2007an}.  

Although other measures over inflationary spacetime can mitigate this divergence, we can imagine it being regulated by including the effect of the collision on the bubble interior. Indeed, for identical bubble collisions, colliding bubbles merge and this regulates the infinity because the collision region has a finite extent in $\xi$, cutting off the exponentially growing of volume at large $\xi$ that is the source of the divergence. Neglecting the effect of overlapping collisions and considering vacuum bubbles, Dahlen~\cite{Dahlen:2008rd} computed the volume fraction inhabited by observation-side instanton-born to overlap-born observers as $f \sim \lambda H_F^{-4} H_F^2/H_I^2$ where $H_F$ is the Hubble constant in the false vacuum, $H_I$ is the Hubble constant in the true vacuum, and $\lambda$ is the nucleation rate per unit four-volume. We expect small corrections to this result due to distortions to the geometry of the reheating surface, but otherwise, our results support this conclusion.

For non-identical bubble collisions, the story is more complicated. The divergence arises from collisions whose boundary is far from the observer, i.e. in overlap-born far-from-boundary observers. As we found above, the only observable effect for such observers is the rescaling of the locally observed spatial curvature; inhomogeneities are largely unobservable. In the case where the post-collision domain wall accelerates out of the observation bubble, it is a good assumption to postulate that the bubble interior is not disturbed, implying that the results of previous work should be largely valid. Although we cannot simulate multiple collisions, it is plausible to imagine that they have a cumulative effect. This would lead to a correlation between the number of collisions and the level of spatial curvature. For some critical number of collisions, inflation in the observation bubble would be completely disrupted, and the curvature would be order one.\footnote{This could be quite sensitive to the inflationary physics inside.  For example, in ``inflection point" small-field models, we expect that inflation could be completely disrupted by even one collision, as suggested in ~\cite{Aguirre:2008wy}.  We leave a more detailed study of different models for later work.} Beyond this, it is unclear that the bubble interior could sustain an arbitrary number of additional collisions without the formation of curvature singularities, or the constant-field hypersurfaces becoming spacelike. In cases where the post-collision domain wall accelerates into the observation bubble, the story should be similar, although the critical number of collisions will be far fewer due to the far larger effect of each collision on the observed curvature. In both cases, it is plausible that the divergent number of collisions is regulated due to back-reaction, as suggested in Ref.~\cite{Aguirre:2009ug}. 

Assessing the probability for observing different levels of curvature is beyond the scope of this paper. The answer could have dramatic importance for the predicted level of curvature from false vacuum eternal inflation, which appears to be largely uncorrelated with the canonical prediction $\Omega_k^{\rm (I)}$ for a measure incorporating volume weighting.

It is interesting in general that infinite spacelike surfaces in eternal inflation appear to be not just generic~\cite{Aguirre:2010rw}, but also more robust than previously expected, as long as the intra-bubble inflation is robust to small perturbations in the field.  For example, an inflationary bubble is safer than previously supposed, as our results show that -- at least for the models we have considered -- even an encroaching bubble would not invade our post-inflationary spacetime and destroy our local universe.\footnote{Though, alas, decay of our vacuum still could.}

\section{Conclusions}

Using a new method for extracting observables from cosmological simulations, we have extended previous work on predicting observables from cosmic bubble collisions in eternal inflation to the entire collision spacetime. This method amounts to a coordinate transformation on comoving hypersurfaces which takes the metric around a point to the spatial section of a perturbed FRW universe in comoving gauge. This allows us to directly extract the local scale factor and comoving curvature perturbation in the neighbourhood of any point. Applying this procedure to an ensemble of points yields predictions for cosmological observables from any vantage point in the post-collision spacetime. 

Our primary goal was to apply this new method to study cosmological observables accessed by overlap-born observers: hypothetical observers who are comoving with respect to the perturbed part of the collision spacetime. The study of instanton-born observers, who are comoving with respect to the un-perturbed part of the collision spacetime, was the subject of previous work. However, the methods employed in this previous work could not be applied to overlap-born observers. 

We studied observables for a single scalar field model of eternal inflation which allows for collisions between identical or non-identical bubbles. In single field models, we can identify the comoving hypersurfaces with surfaces of constant scalar field. Identical bubbles merge when they collide, forming smooth  spacelike constant field hypersurfaces at late times  spanning the interior of both bubbles. Contrary to previous assumptions in the literature~\cite{Aguirre:2008wy,Aguirre:2009ug,Chang_Kleban_Levi:2009,Czech:2010rg,Gobbetti_Kleban:2012,Kleban_Levi_Sigurdson:2011,Kleban:2011pg,Feeney_etal:2010dd,Feeney_etal:2010jj}, instanton- and overlap-born observers near the collision boundary do not see the same cosmological signature. Overlap-born observers near the collision boundary experience a curvature perturbation that is everywhere non-zero in their neighbourhood. Moving deeper into the collision region, overlap-born observers in collisions between identical bubbles would experience an observable universe containing a very nearly quadratic planar comoving curvature perturbation. The leading observable in this case would be a CMB quadrupole and a spatially varying negative curvature. 

The collision between non-identical bubbles is qualitatively different, since the interiors of the colliding bubbles are separated by a domain wall produced in the collision. This domain wall can accelerate into or out of the observation bubble, depending on the structure of the scalar field potential. Rather remarkably, after a few $e$-folds of inflation, the surfaces of constant field re-arrange themselves to be everywhere spacelike, although the geometry of these hypersurfaces is quite different from what it would have been in the absence of a collision. Moving deep into the collision region, the universe accessible to overlap-born observers becomes increasingly homogeneous. Therefore, in contrast to the collision between identical bubbles, when non-identical bubbles collide, there is an infinite class of overlap-born observers who would not have access to any observables indicating that there was a bubble collision in their past. Only those observers in the neighbourhood of the collision boundary would have any hope of detecting traces of the collision. 

The internal degrees of freedom of the post-collision domain wall produced in the collision between non-identical bubbles can be excited by the collision, producing an additional oscillatory contribution to the curvature perturbation. Some overlap-born observers have access to these wall modes, which would contribute an aligned set of contributions to their low-order CMB multipoles.

Our results suggest a few possible modifications of the observational search strategy for cosmic bubble collisions. Clearly it does not make sense to treat the overlap-born observers in the neighbourhood of the collision boundary the same as the instanton-born observers. To a good approximation, such overlap-born observers experience a matched linear and quadratic comoving curvature perturbation. Deeper into the collision region, where overlap-born observers have access to a smooth comoving curvature perturbation varying on long wavelengths, the observable signature is not as distinctive. In this case, it is possible that an observation of negative spatial curvature could rule out some models over some range of initial bubble separations. However, it is clear that many models could be consistent with any level of observed negative spatial curvature and the observed CMB quadrupole. It may however be possible to include CMB polarization and large scale structure to determine the level of planarity of the large-scale curvature perturbation in our universe. This could provide suggestive, although certainly not definitive, evidence that we could be a overlap-born observer in a collision spacetime. 

We also considered how our results may affect the probability for observing various bubble collsion signatures. Previous work identified a divergence in the predicted number of collisions for most observers, nearly all of whom would be of the overlap-born far-from-boundary type. As we found in this paper, overlap-born far-from-boundary observers inhabit a homogeneous universe with a re-scaled spatial curvature. Considering the cumulative effect of multiple collisions, it is plausible to conjecture that the divergence is regulated by back-reaction, and that multiple collisions act to further re-scale the curvature. Important work remains to be done computing the measure over observed curvature implied by this result. Work remains to be done on incorporating multiple scalar fields as well, which undoubtedly have a far richer phenomenology. 

\acknowledgments 

This work was partially supported by a New Frontiers in Astronomy and Cosmology grant \#37426. Research at Perimeter Institute is supported by the Government of Canada through Industry Canada and by the Province of Ontario through the Ministry of Research and Innovation. This work was supported in part by National Science Foundation Grant No. PHYS-1066293 and the hospitality of the Aspen Center for Physics. MCJ is supported by the National Science and Engineering Research Council through a Discovery grant. AA was supported in part by a time release grant from the Foundational Questions Institute (FQXi), of which he is Associate Director. HVP was partially supported by the European Research Council under the European Community's Seventh Framework Programme (FP7/2007- 2013) / ERC grant agreement no 306478-CosmicDawn. 
\bibliographystyle{JHEP}
\bibliography{largerthansky}

\end{document}